\documentclass[twocolumn,floatfix,superscriptaddress,longbibliography,nofootinbib]{revtex4-1}
\RequirePackage[sort&compress]{natbib}

\usepackage{color}
\usepackage{amssymb,amsfonts,amsmath}
\usepackage{bm}
\usepackage[pdftex]{graphicx}
\usepackage[dvipsnames*,svgnames]{xcolor}
\usepackage{comment}

\usepackage[colorlinks = true,
            linkcolor = blue,
            urlcolor  = blue,
            citecolor = blue,
            anchorcolor = blue]{hyperref}

\usepackage{lipsum}

\usepackage{dcolumn}
\usepackage{bm}

\usepackage{subcaption}
\usepackage{xcolor}

\newcommand{\heading}[1]{{\vspace{0.25truecm}\noindent\textbf{#1.}}}

\usepackage{xcolor}
\definecolor{RoyalBlue}{HTML}{4169e1}
\definecolor{ForestGreen}{HTML}{228b22}
\definecolor{Aquamarine}{HTML}{00B5BE}
\colorlet{TransAquamarine}{Aquamarine!30!white} 
\definecolor{RedViolet}{HTML}{A1246B}
\definecolor{RedOrange}{HTML}{F26035}
\colorlet{TransRedOrange}{RedOrange!30!white} 
\definecolor{lightgray}{gray}{0.9}

\usepackage{setspace}
\usepackage{comment}

\usepackage{mdframed}
\usepackage{enumitem,amssymb}
\newlist{todolist}{itemize}{2}
\setlist[todolist]{label=$\square$}
\usepackage{pifont}

\usepackage{todonotes}

\newcommand{\rev}[1]{{\color{black} #1}}

\usepackage{tcolorbox}
\tcbuselibrary{skins}

\usepackage{xcolor}
\definecolor{darkred}{HTML}{A10015}

\newtcolorbox{mybox}[1][]{%
  enhanced,
  colback=darkred!10,             
  colbacktitle=darkred,        
  coltitle=white,              
  colframe=darkred!50,           
  boxrule=1pt,
  arc=2mm,
  left=2mm,
  right=2mm,
  top=1mm,
  bottom=1mm,
  width=\textwidth,            
  fontupper=\fontsize{10}{11}\selectfont,
  before skip=10pt,
  after skip=10pt,
  title=#1,                    
  boxed title style={
    enhanced,
    colback=gray!40,
    colframe=black!50,
    boxrule=0.8pt,
    arc=1mm,
  },
}

\usepackage{verbatim}
\usepackage{lineno}

\usepackage{draftwatermark}

\begin{document}

\title{Decoding the Architecture of Living Systems}
\author{Manlio De Domenico}\thanks{Corresponding author: \url{manlio.dedomenico@unipd.it}}
\affiliation{Department of Physics and Astronomy "Galileo Galilei", University of Padua, Via F. Marzolo 8, 315126 Padova, Italy}
\affiliation{Padova Neuroscience Center, University of Padua, Via G. Orus 2, 35131 Padova, Italy}
\affiliation{Padua Center for Network Medicine, University of Padua, Via F. Marzolo 8, 315126 Padova, Italy}
\affiliation{Istituto Nazionale di Fisica Nucleare, Sez. Padova, Italy}
\date{\today}

\begin{abstract}
The possibility that evolutionary forces -- together with a few fundamental factors such as thermodynamic constraints, specific computational features enabling information processing, and ecological processes -- might constrain the logic of living systems is tantalizing. However, it is often overlooked that any practical implementation of such a logic requires complementary circuitry that, in biological systems, happens through complex networks of genetic regulation, metabolic reactions, cellular signalling, communication, social and eusocial non-trivial organization. Here, we review and discuss how circuitries are not merely passive structures, but active agents of change that, by means of hierarchical and modular organization, are able to enhance and catalyze the evolution of evolvability. By analyzing the role of non-trivial topologies in major evolutionary transitions under the lens of statistical physics and nonlinear dynamics, we show that biological innovations are strictly related to circuitry and its deviation from trivial structures and (thermo)dynamic equilibria. 

We argue that sparse heterogeneous networks such as hierarchical modular, which are ubiquitously observed in nature, are favored in terms of the trade-off between energetic costs for redundancy, error-correction and mantainance. We identify three main features -- namely, interconnectivity, plasticity and interdependency -- pointing towards a unifying framework for modeling the phenomenology, discussing them in terms of dynamical systems theory, non-equilibrium thermodynamics and evolutionary dynamics. Within this unified picture, we also show that “slow” evolutionary dynamics is an emergent phenomenon governed by the replicator-mutator equation as the direct consequence of a constrained variational nonequilibrium process. Overall, this work highlights how dynamical systems theory and nonequilibrium thermodynamics provide powerful analytical techniques to study biological complexity. 
\end{abstract}

\maketitle

\tableofcontents

\section{Introduction}

\subsubsection{From systems to living systems}

A system is a collection of interrelated units that interact and process matter, energy, and information~\cite{miller1976nature}. A living system is a special type of open system characterized by self-regulation and hierarchical \rev{organization}: \rev{it maintains a dynamic internal balance while continuously interacting with its environment, adapting to external changes~\cite{rosen1958relational1,rosen1959relational2,miller1976nature}.}

\rev{Here, the term “structure” refers to the spatial arrangement of components at a given time, providing a snapshot of its organization. The terms “topology” and “network” indicate the general configuration of connections and relationships among these components, irrespective of their physical location. A “process” denotes any time-dependent transformation of matter, energy or information, while “behavior” designates the system's observable responses over time, arising from internal processes and environmental coupling.}

\rev{Classical mechanistic approaches -- often exemplified by reductionist Newtonian reasoning -- seek to understand systems by decomposing them into constituent parts. However, they fail when applied to living systems, whose essential properties emerge from complex interactions among components~\cite{ulanowicz1999life}. Similarly, strictly adaptationist views grounded in Darwinian evolution emphasize natural selection as the main driver of change but overlook the spontaneous self-organization and feedback dynamics inherent to complex biological processes~\cite{morowitz1993beginnings}.}

\rev{Living systems therefore cannot be fully understood through deterministic mechanics or adaptationist approaches alone: their organization emerges from relational processes that integrate structure and function beyond linear cause--effect chains~\cite{rosen1958relational1,rosen1959relational2,ulanowicz1999life}. Understanding these relational principles provides the foundation for the framework developed in the following sections. To illustrate why structure, topology, and process must be treated jointly, we present in Box A a quantitative example showing how informational and energetic constraints shape the organization of biological networks, from yeast to the human brain.}

\subsubsection{From emergence to architecture}

Understanding the interplay between multiple components of a complex system~\cite{gell2002complexity} is crucial for unraveling its behavior across various scales and conditions. To this aim, one requires connecting their informational logic with the topological organization that implements it. Building on an broad corpus of studies at the crossroad of evolutionary biology, systems biology, statistical physics, computer science, ecology and social sciences, it has been recently conjectured that it is possible to consider a few constraints of thermodynamical, informational and ecological origin, together with evolutionary forces, to constrain the logic of living systems~\cite{sole2024fundamental}. 

\begin{figure*}[!t]
\centering
\includegraphics[width=\textwidth]{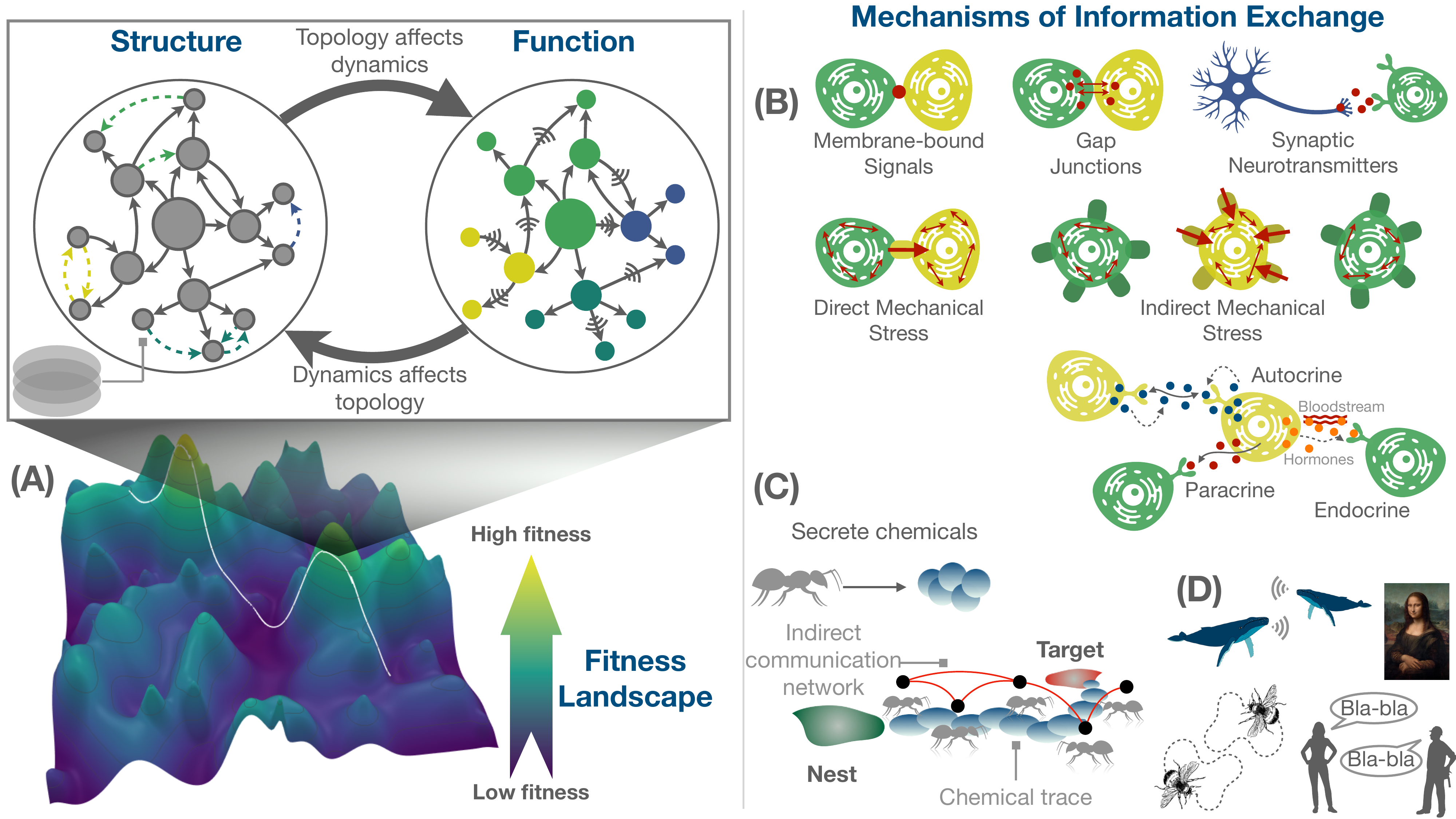}
\caption{\label{fig:mechanism}\textbf{Complex adaptive networks}. (A) The dynamic interplay between structure and function in complex adaptive networks, governed by a fitness landscape, highlights how network topology influences dynamics and vice versa, shaping the network's plasticity and its evolutionary trajectory. \rev{Each possible network configuration corresponds to a point in the fitness landscape, whose elevation reflects functional performance -- e.g., efficiency, robustness or information throughput -- under given constraints. Adaptive dynamics describe how mutations or rewiring events modify topology, enabling the system to explore and climb this landscape toward locally or globally optimal architectures.} (panel readapted from~\cite{berner2023adaptive}) (B) Intercellular interactions unfold through different mechanisms. Contact-based exchange includes signaling via cell surfaces, connecting through structures like gap junctions to directly transfer signals, or exerting direct mechanical stress. Mechanisms based on soluble factors involve cells releasing neurotransmitters across synapses, influencing themselves or nearby cells (autocrine and paracrine signaling), or sending hormones through the bloodstream to distant cells (endocrine signaling), along with creating changes in their environment that indirectly apply mechanical stress to nearby cells (panel readapted from~\cite{yang2021engineered}). (C) In eusocial systems like ant and termite colonies, indirect communication occurs through stigmergy, a phenomenon where each insect's secretion of chemicals onto a physical substrate leads to environmental changes that allow for self-coordinating the collective activity of the colony~\cite{theraulaz1999brief,holland1999stigmergy}. (D) Visual signals are encoded in the waggle dance of bees, for social learning and signaling information about food sources~\cite{dong2023social}. Auditory communication is observed in sperm whales, who use codas to maintain social interactions~\cite{sharma2024contextual}. In humans, direct communication is facilitated through spoken language (see~\cite{fedorenko2024language} and refs. therein), \rev{as well as visual cues and possibly chemical signals}, 
while indirect and asynchronous communication occurs through various art forms, enabling the expression and transmission of complex cultural and emotional information across individuals and generations.}
\end{figure*}

However, while the \emph{logic} used by formal systems can perfectly exist as an abstraction, serving specific computational or algorithmic functions, living systems represent a complex interplay of physical and biochemical processes that cannot be fully understood through abstraction alone. \rev{In this context, \emph{logic} denotes the set of causal and informational rules that map inputs to outputs and govern the behavior of the system -- its internal operations and constraints—whereas \emph{circuitry} refers to the material or biochemical realization of those rules through interacting components and connections. Concretely, the same logical organization can be implemented by different circuitries: in genetic systems, by transcriptional and post-transcriptional interactions among genes and proteins; in metabolism, by chains of enzymatic reactions linking metabolites; in neural or ecological systems, by physical or functional connections mediating information flow (see Fig.~\ref{fig:mechanism} and \ref{fig:empirical}).}

In fact, living systems embody some type of logic and circuitry implemented through the intricate, often non-linear interactions of a large number of biochemical and biological components, \rev{whose informational organization can be interpreted through the lens of algorithmic structure and reprogrammability~\cite{zenil2012life,zenil2019algorithmic}}. These components form networks that dynamically respond to both internal and external stimuli, driving the multifaceted processes characteristic of life, such as adaptation~\citep{nagelkerken2020trophic} and self-regulation~\citep{alam2017self}, while subject to thermodynamical, informational and ecological limits that shape their functionality. 

\rev{These constraints, however, are not sufficient to determine the architecture of living systems: not all circuitries support behaviors such as stability-while-diverse, robustness, modularity, or hierarchy, impossible in trivial topologies such as linear chains, trees, cliques or even simple random networks.}
Conversely, such concepts are better understood in terms of heterogeneous random networks (e.g., with fat-tailed connectivity distribution)~\cite{barabasi1999emergence,caldarelli2002scale,serafino2021true} with mesoscale organization perfectly captured by stochastic block models~\cite{peixoto2014hierarchical}. Accordingly, defining the logic alone without specifying the intervening networks would only provide a partially picture to grasp the complexity of living systems. 

\subsubsection{Toward a blueprint for living systems?}

\rev{Despite the abundance of molecular and systems-level data, we still lack a quantitative framework that links microscopic biochemical interactions to macroscopic adaptive behavior. Existing models often capture either the dynamical or the structural aspect of living systems, but rarely their mutual interdependence which unfolds across multiple layers and levels of organization. This gap motivates the use of network theory as an integrative language, able to provide quantitative descriptors of structure-function coupling across biological scales.} 

On the one hand, this challenges the creation of a comprehensive “theory of everything”~\cite{laughlin2000theory} in biological terms, even if only weak emergence is assumed, as the physical “implementation” of these systems involves layers of complexity that defy simple reductionism~\cite{anderson1972more,laughlin2000middle,artime2022origin}. Note that we do not want to enter into the debate of weak versus strong emergence here, primarily because there is insufficient theoretical and experimental evidence to classify biological processes as fundamentally irreducible or reducible to lower-level laws. Emergent phenomena are undeniable in the biological sciences and crucial for understanding complex processes. For instance, pattern formation in developmental biology involves complex interactions and feedback loops that cannot be fully explained by examining individual cellular behaviors alone~\cite{turing1952chemical,kondo2010reaction}.

Similarly, cooperation in socio-ecological systems, which includes both competitive and collaborative interactions among various species, illustrates how collective behaviors emerge that are not predictable from the properties of individual organisms~\cite{levin1998ecosystems,nowak2006five}. Accordingly, theories that are limited to the lowest level of description of our reality are not sufficient to capture these multifaceted processes that characterize living systems. Therefore, proposing a hypothetical theory of everything, even under assumptions of only weak emergence, might be extremely challenging. This is because even weakly emergent properties, though in principle reducible to simpler laws, often involve complexities and interactions at scales that defy straightforward reduction, making a complete analytical theory elusive.

On the other hand, developing modular and scalable models that can be integrated as needed to describe different levels of biological complexity can lead to a blueprint for the architecture of living system that is still an appealing, desirable and possibly achievable goal~\cite{simon1962architecture}.
By combining a few fundamental processes -- recurring across different spatial and temporal scales -- governing the function of living systems (stability, diversity, robustness, resilience, ...) with constraints -- accumulating across those scales -- that pose fundamental limits to such processes, it is possible to avoid the pitfalls of a theory of everything for living systems, while building an operational framework for their understanding. The advantages could significantly outweigh the costs, offering a framework that could guide the development of new therapeutic strategies and interventions in ecosystem management.

\onecolumngrid
\begin{mybox}[\textbf{Box A:} Constraints and Generative Principles of Living Systems]
\hspace{0.3truecm}To better understand why the interplay between structure, topology and processes involving the exchange of energy, matter and information is relevant, let us consider some practical examples. 

\vspace{0.1truecm}\hspace{0.3truecm}There are nearly 6,000 genes in \emph{Saccharomyces cerevisiae}, the budding yeast, involving about 170,000 pairwise interactions~\cite{costanzo2010genetic}, and there are about 86 billions neurons in the human brain~\cite{azevedo2009equal}, with about a trillion synapses per cubic centimeter of cortex~\cite{drachman2005we}. If interactions, and their organization, enable novel functions that -- in principle -- can allow the budding yeast to potentially adapt to any uncertain environment or the human brain to achieve novel forms of intelligence or an ecosystem to benefit from an uncountable number of mutualistic relationships, what is preventing those systems to unlock their full potential?

\vspace{0.1truecm}\hspace{0.3truecm}Actually, assuming that such a full potential relies only on building more connections or adding a larger number of interacting units is naive, since it is not accounting for the energetic costs for building, error-correcting~\cite{hopfield1974kinetic,bennett1979dissipation,banerjee2017elucidating}, repairing~\cite{bekker2010assembly,ciccia2010dna,polo2011dynamics} and maintaining~\cite{hoeijmakers2001genome,wallace2010bioenergetics,lane2010energetics,navarrete2011energetics} such biological systems, factors that impose constraints that are effectively reflected by non-trivial connectivity patterns. For instance, let us consider again the human brain and its huge number of connections, nearly $10^{14}$, among approximately $10^{11}$ units. If evolution had to act as an engineer (and, likely, it did not~\cite{jacob1977evolution}) it should associate to the presence or absence of a connection at least one bit, like in a modern memory storage: 0 for a missing link and 1 for a present link. 

\hspace{0.3truecm}Under the simplifying assumption of independence, the microstate space describing all the possible interactions has size $2^{10^{22}}$ and, accordingly, the capacity of an ideal “device” to losslessly store this information is the maximum Shannon entropy $\log_{2} N$, that in this case reduces to $10^{22}$~bits (about 1~ZB), which is astronomically large and thus energetically expensive for computational purposes (see~\cite{bennett1982thermodynamics} and Refs. therein). In fact, this kind of device is DNA, and for humans it consists of only 3 billion base pairs, with a microstate space of size $4^{3\cdot 10^{9}}$ and a corresponding capacity of $6\cdot 10^{9}$~bits (roughly 0.7~GB), which is several orders of magnitude smaller than what would be needed for our storage purpose. In computer science, however, there are more efficient ways to reproduce such a network: one might store it as an adjacency list. Using a 64 bit code to represent unsigned integers, the microstate space has size of about $10^{19}$ and can be used to store only the information about the neighbors of each unit, spending “only” $1.3\cdot 10^{16}$~bits (about 1.4~PB): still huge, but far less than the previous astronomical estimate but still six orders of magnitude larger than what DNA can achieve.

\vspace{0.1truecm}\hspace{0.3truecm}Still, evolution overcame this necessity of specifying exactly which neuron connects to other neurons: they are not allowed to explore the whole microstate space. Instead, it is cheaper to store the “blueprint” to build this gigantic neural structure by encoding the fundamental mechanisms governing each individual neuron for creating connections. This approach introduces some level of stochasticity, so the replication cannot be perfect, but the implementation of control and error-correction mechanisms can satisfactorily solve the problem~\cite{banerjee2017elucidating}. In fact, instead of storing and replicating the network information, the generative mechanisms are transmitted instead: provided that control mechanisms reduce the probability of errors, this solution is far more elegant and, at the same time, more efficient. Ongoing research is exploring potential methods to implement such generative mechanisms through evolutionary learning~\cite{pezzulo2016top,vanchurin2022toward,bentley2022evolving,levin2023darwin,mitchell2024genomic}. \rev{However, such generative programs do not uniquely determine the resulting connectivity. Developmental noise, stochastic gene expression and environmental influences introduce variability that cannot be predicted from the genome alone. Consequently, brain architecture should be regarded as the outcome of both genetically encoded constraints and non-deterministic developmental processes modulated by epigenetic and environmental factors~\cite{rakic2009evolution,holtmaat2009experience}.}
We have used the human brain as a vivid example, but the same argument can be developed to describe the emergence of regulatory and metabolic networks governing cellular processes. 
\end{mybox}
\twocolumngrid

Using the same analogy as before, this goal requires to combine, rather than separate, the intervening logic with the relevant circuitry. At first sight, it might be plausible to think that such a blueprint should account for as much as possible microscopic information about the fundamental constituents of a living system and their specific processes, across all possible intervening spatial and temporal scales. However, this might not be the case given the evidence that it is possible to describe collective dynamics of some components at a given scale by coarse-graining such systems~\cite{yang2021engineered,theraulaz1999brief,holland1999stigmergy,meunier2010modular,wild2021social,dong2023social,sharma2024contextual}, while neglecting the majority of overwhelming details at lower scales. In fact, such collective dynamics are often more predictive of the future behavior of a system than the information available at individual level. This is almost certainly due to the spontaneous appearance of novel intervening mechanisms at the higher levels of description~\cite{laughlin2000middle}, which is ubiquitous from cells to societies~\cite{artime2022origin}.

Here, we propose that by considering living systems as adaptive networks of networks (i.e., multilayer systems), it is possible to map the evolution of each unit's state not only by internal dynamics and pairwise interactions but also by external drivers and environmental factors. This framework employs a non-autonomous system of coupled ordinary differential equations (ODEs) to capture the continuous influence of these variables on the living system’s internal and external structures and its subsequent feedback on the dynamics, \rev{consistent with eco-evolutionary feedback models~\cite{tilman2020evolutionary}}. By distinguishing between external inputs (e.g., targeted interventions) and environmental states (e.g., broader changes affecting a system as a whole, not necessarily targeted), the framework provides a granular understanding of how specific interventions can be differentiated from broader environmental changes, offering an operational way to model living systems as complex networks with adaptive interactions and co-evolutionary dynamics. Adaptation unfolds through two mechanisms: plasticity, allowing functional units and sub-systems at play to change their structural organization to meet the current requirements, and information exchange, allowing to propagate perturbations (e.g., useful to enable specific functional or behavioral responses, tasks or defense) and communication. Figure~\ref{fig:mechanism} shows how the concept of complex adaptive networks~\cite{berner2023adaptive} can be used to model the interplay between structure and function, as well as a non-comprehensive set of mechanisms of information exchange from cells to eusocial and social systems.

\textit{Organization of this study. --} Section~\ref{sec:basic} will provide an overview of the fundamental concepts about complex networks which are relevant for the subsequent sections: network ensembles, operators, closed cycles, the emergence of hierarchical organization and the generalized thermodynamics built from network density matrices. \rev{These concepts are essential to clarify how perturbations propagate, robustness and adaptability coexist, and how network structure constrains or enables functional change across scales.} In Sec.~\ref{sec:major_evol_transitions} we will first identify the major evolutionary transitions~\cite{szathmary1995major,szathmary2015toward} describing -- at different scales and with increasing complexity -- the emergent life forms as we know them, while focusing on the appearance and role of networks to characterize each transition. 
We will consider distinct levels of organization in biological systems, from molecular to social and ecological scales, characterizing each level in terms of the intervening networks from a phenomenological perspective, to highlight the desirable features to be reproduced by a suitable modeling framework. In Sec.~\ref{sec:architecture} we will summarize into a comprehensive scheme the main observational properties of living systems and propose three fundamental features to characterize their architecture in the most general terms, while in Sec.~\ref{sec:model} we will formalize such insights into a multiscale model. It is important to clarify that this model is not intended as a universal theory that explains all dynamics of living systems, but rather as a flexible framework designed to guide the understanding and description of complex biological architectures from the perspective of physics. Like the principle of least action in classical mechanics, which provides a systematic approach to deriving equations of motion without defining specific outcomes, our framework offers a structured methodology to explore the diverse and adaptive behaviors of living systems through a mathematical framework encompassing a broad variety of observations. This approach acknowledges the intrinsic variability and complexity of biological process, \rev{while recognizing that no model can capture all possible configurations and interactions in such systems}: in fact, we will talk about meaningful mathematical representations and sets of constraints in general terms, without referring to specific ones. In the Discussion, we will reconcile this framework with the concept of evolvability, which is fundamental to characterize living systems.


\section{Principles of complex networks}\label{sec:basic}

Complex networks provide a versatile framework for describing interactions and structural organization in diverse systems~\cite{albert2002statistical,newman2003structure,boccaletti2006complex,cimini2019statistical,boguna2021network,artime2024robustness}. Mathematically, a network can be represented by means of an adjacency matrix $\mathbf{A}$, where $A_{ij}=1$ if units $i$ and $j$ are connected, and $A_{ij}=0$ otherwise. Note that when dealing with empirical networks, missing or spurious connections, as well as measurement biases and other factors might require an adequate inferential process to reconstruct the underlying structures~\cite{peel2022statistical}.

A key feature of these networks is their connectivity, where $k$ denotes the degree of a single unit measured as the number of its adjacent edges, e.g., $k_{i}$ indicates the degree of unit $i$. By indicating with $\mathbf{D}$ the diagonal matrix with entries $D_{ii}=k_{i}$, it is possible to represent the network by means of another operator, the combinatorial Laplacian matrix defined by $\mathbf{L}=\mathbf{D}-\mathbf{A}$. The Laplacian matrix is important because its eigenvalue spectrum regulates a variety of dynamical processes, from diffusion~\cite{masuda2017random} to synchronization~\cite{arenas2006synchronization}.

The degree distribution $P(k)$, with its moments $\langle k^{n}\rangle=\sum_k P(k)k^{n}$, plays a fundamental role in determining the overall connectivity, characterizing a wide variety of topological features, from percolation to robustness to internal failures or external disturbances~\cite{newman2003structure,dorogovtsev2008critical}. 

Connectivity is directly related to network density. Let us consider a simple unweighted and undirected complex network of size $N$ and average degree $\langle k \rangle$. The edges density $\rho$ of the network is defined by the ratio between the total number of edges $2m$ in the network and the maximum possible number of edges, i.e., $\binom{N}{2}$. Since the minimum number of edges in a connected network is $N$ and the maximum possible number of edges scales with $N^2$, it is possible to define a scaling ansatz to characterize network sparsity: let us assume that $2m \sim N^{\beta}$, with $1\leq \beta \leq 2$ \rev{being a dimensionless scaling exponent}. Accordingly, for large networks ($N \gg 1$):
\begin{eqnarray}
\rho = \frac{2m}{N(N-1)} = \frac{\langle k \rangle}{N} \sim N^{\beta-2}, 
\end{eqnarray}
or equivalently $\langle k \rangle \sim N^{\beta -1}$. If the edges density is constant with respect to system size $N$, the network is dense, otherwise it is defined to be sparse, which is the case observed in real-world systems~\cite{ghavasieh2024diversity,busiello2017explorability}. Similarly, it is possible to define sparsity in terms of the scaling of the average degree.

\subsection{Closed cycles in random networks}\label{sec:cycles}

For the subsequent analysis, it might be beneficial to introduce here the concept of network cycles. Let $k_i$ denote the degree of a node $i$, in a simple unweighted and undirected network of size $N$. Let us consider the network ensemble known as the configuration model, which is fully specified by the degree sequence $k_{1}, k_{2}, ..., k_{N}$~\cite{molloy1995critical}. In this network ensemble, the probability $p_{ij}$ that a link exists between nodes $i$ and $j$ is approximated as $p_{ij} = \frac{k_i k_j}{2m}$, where $2m = \sum\limits_i k_i$ is the total number of edges.

A rough estimation of the probability $p(n)$ of closed simple cycles of length $n$, formed by any sequence of nodes $j_1, j_2, \ldots, j_n$, can be obtained \rev{by starting from the expected count of length-$n$ simple cycles\footnote{\rev{Note that there is a pre-factor $1/2n$ to accounts for the two choices of orientation and the $n$ choices of starting point on the cycle.}}}, as
\begin{eqnarray}
\sum\limits_{j_1, \ldots, j_n = 1}^{N} \left(\frac{k_{j_1} k_{j_2}}{2m}\right)  \ldots \left(\frac{(k_{j_n}-1) (k_{j_1}-1)}{2m}\right) = \left(\kappa-1\right)^n,\nonumber
\end{eqnarray}
where $\kappa=\frac{\langle k^2 \rangle}{\langle k \rangle}$ is often referred to as the Molloy-Reed parameter or, equivalently, as the heterogeneity coefficient, playing an important role in percolation transitions (see later).

This expression highlights the dependency of cycle formation on the first two moments of the degree distribution, reflecting the importance of network's heterogeneity in connectivity. 

For instance, assuming an homogenous degree distribution such as an Erd\H{o}s-R\'enyi network, characterized by a constant probability of link existence between any pair of nodes, we have that $\langle k^2 \rangle = \langle k \rangle(\langle k \rangle+1)$ and $p(n)\sim \langle k\rangle^{n}$. Since $\langle k\rangle\sim N^{\beta-1}$, we have $p(n)\sim N^{(\beta-1)n}$ from which we see that there is an important difference between sparse ($\beta \simeq 1$) and dense ($\beta \simeq 2$) homogenous networks in terms of cycles: $p_{\text{dense}}(n)/p_{\text{sparse}} \sim N^{n}$, showing an exponential increase in cycles between the two regimes.

Similar arguments, assuming a power-law degree distribution $P(k)\sim k^{-\gamma}$, lead to a more complicated form. However, by considering that heterogeneous networks are characterized by a natural cutoff~\cite{cohen2000resilience,dorogovtsev2001size} $k_{\max}\sim N^{1/(\gamma-1)}$, and that most empirical networks of biological interest are characterized by $2 < \gamma < 3$, using $\gamma=2.5$ we obtain a scaling as $N^{n/3}$ (\rev{or $N^{n/4}$ by using the structural cutoff for uncorrelated simple networks}), which is much larger than the one of a sparse homogeneous network and smaller than the one of a dense homogeneous network.

\subsection{Emergence of hierarchy in random networks}\label{sec:hierarchy}

Since most empirical networks of biological interest are hierarchically organized~\cite{ravasz2002hierarchical,ravasz2003hierarchical,yook2004functional,corominas2013origins}, it is plausible to wonder how hierarchical networks can emerge. On the one hand, they can be the result of some optimization process~\cite{carlson2002complexity,doyle2005robust} that requires a well-defined target and some cost function to be optimized. On the other hand, we wonder if they can be the result of some stochastic process that, at equilibrium, can be described by an adequate block model~\citep{peixoto2014hierarchical}.

\begin{figure*}[!t]
\centering
\includegraphics[width=\textwidth]{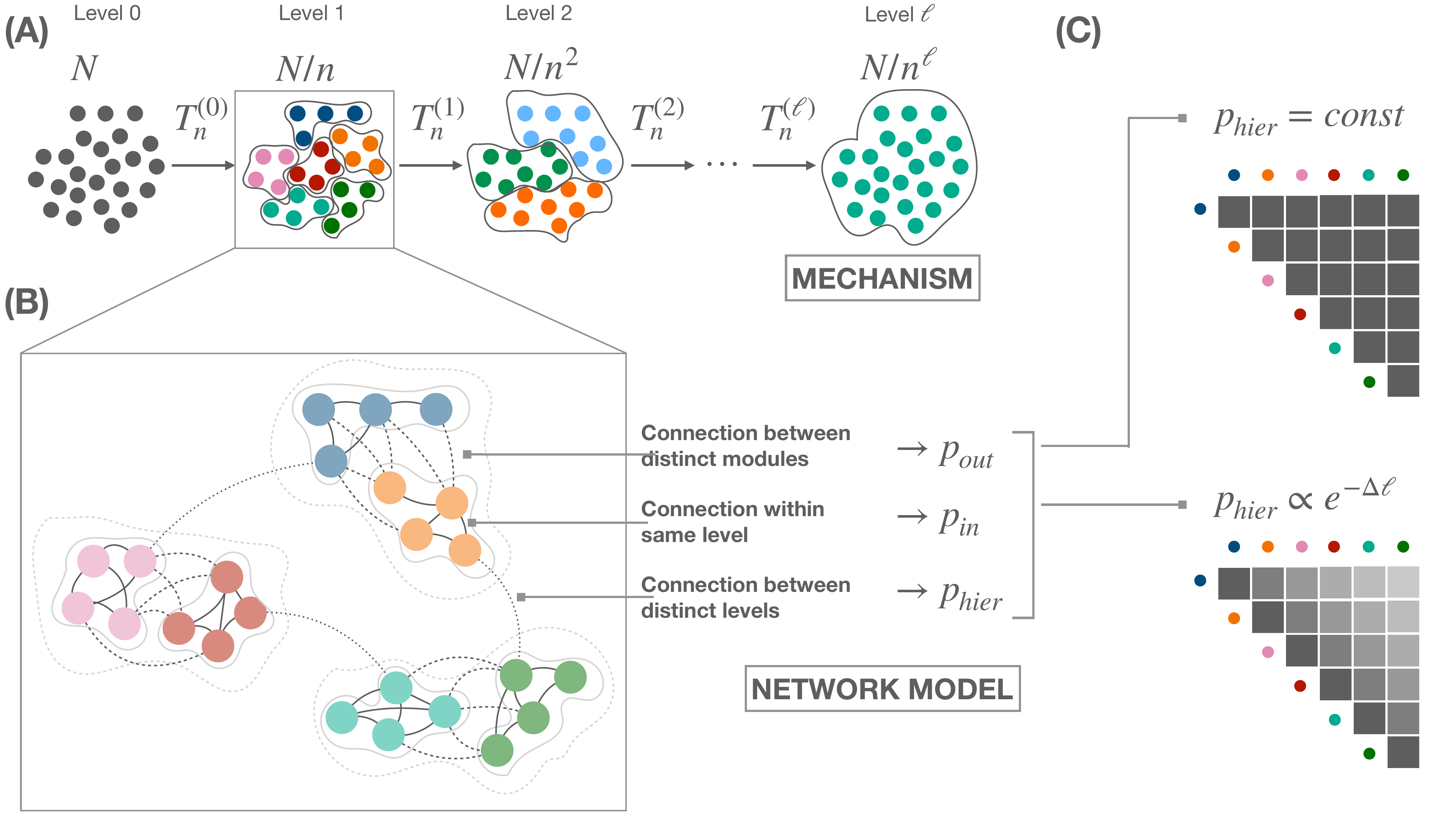}
\caption{\label{fig:hierarchy}\textbf{Formation of a hierarchical structure.}
(\textbf{A}) Illustration of a toy model where $N$ elementary units, initially disconnected, interact to form assemblies of size $n < N$. The resulting groups, now $N/n$, in turn aggregate into new assemblies of size $n$, and so on, until the process ends for some value $\ell = \log_{n} N$ that sets the depth of the hierarchy. The transition between two consecutive levels of the hierarchy is denoted by $T_{n}^{(i)}$.
(\textbf{B}) Schematic representation of the resulting network model: connections within the same module occur with probability $p_{\text{in}}$, connections between different modules at the same level with $p_{\text{out}}$, and connections across different hierarchical levels with $p_{\text{hier}}$.
(\textbf{C}) Probability matrix corresponding to a three-level hierarchy, where the off-diagonal blocks encode cross-level links. Depending on the mechanism, $p_{\text{hier}}$ can be a constant value or decay exponentially (e.g., $p_{\text{hier}} \propto e^{-\Delta \ell}$).
}
\end{figure*}

To this aim, inspired by the famous parable of the two watchmakers proposed by Herbert Simon~\cite{simon1962architecture,simon1977organization}, let us consider a set of $N$ elementary units and some simple iterative and aggregative process where all pairs of units are allowed to interact and form groups (or components or modules) of $n$ elementary units (see Fig.~\ref{fig:hierarchy}). 

Similarly, Dawkins' concept of the “blind watchmaker” illustrates that complex biological structures emerge from a series of small, undirected modifications, each a random trial that, when selected for functionality, cumulatively builds intricate systems. Just as Simon's modular watchmaker assembles parts in a robust, iterative fashion, natural selection works without foresight, incrementally favoring beneficial variations to eventually yield hierarchical organization~\cite{dawkins1989evolution}.

Accordingly, we can assume in our toy model that the formation of components is supposed to be functional to some task relevant for the survival of the corresponding system. Accordingly, since we do not know \emph{a priori} which configuration of sub-systems is functional, we can further assume that an evolutionary process leading to the form of a given system is allowed to generate, in principle, any sub-system $G_{n,\beta}$ of size $n$ with some probability $P_{n,\beta}(G_{n})$, where $\beta$ is the same exponent appearing in the scaling ansatz for network sparsity. We will use more compactly the notation $P_{n,\beta}$, unless in case of ambiguity, while we explicitly keep the dependence on $\beta$ to gain insights about the results as a function of network density. Each sub-system $G_{n,\beta}$ is “tested” and its probability to be functional is denoted by $\epsilon_{n,\beta}$.

Assuming that the connection of two elementary units in any sub-system $G_{n,\beta}$ in the space of all possible configurations $\mathcal{G}_{n,\beta}$ is equally probable, simple and undirected with probability $p$, and having no information about any predetermined degree sequence, we can use the Erd\H{o}s-R\'enyi canonical ensemble $\mathcal{G}(n,p)$. Accordingly, the probability of sampling a component with exactly $2m=cn^{\beta}\equiv m_{n,\beta}$ edges -- that follows the aforementioned scaling ansatz -- from this ensemble is given by
\begin{eqnarray}
\text{Pr}(m_{n})\equiv P_{n,\beta}=\binom{\binom{n}{2}}{m_{n,\beta}} p^{m_{n,\beta}} (1-p)^{\binom{n}{2} - m_{n,\beta}}.
\end{eqnarray}
We can use this result to calculate the average time $T_{n,\beta}$ needed to sample a functional module of size $n$ and $m_{n,\beta}$ edges. \rev{If $\Delta t_{n,\beta}$ indicates the time required} to build $G_{n,\beta}$, and we assume that each configuration is tested for functionality following a simple yes/no decision process (e.g., performed by natural selection), than we are dealing with a Bernoulli process and 
\begin{eqnarray}
T_{n,\beta}=\frac{1}{P_{n,\beta}\epsilon_{n,\beta}}\Delta t_{n,\beta}
\end{eqnarray}
\rev{provides the average waiting time in an evolutionary search process.} Here, we are considering a rather general case where both the probability of being functional and the time needed to “test” one configuration depend on the edges density of the sub-system.

In the following, let us consider $n\gg 1$, typically 30 or larger. Accordingly, $p\approx 2cn^{\beta-2}$ is the edges density. Since $\beta$ is our free parameter, we are interested in understanding what is the difference in terms of sampling probability between the sparse ($\beta=1$) and the dense ($\beta=2$) regimes. Note that in this case the prefactor $c$ should not be necessarily the same in the two regimes: while $c \leq \frac{1}{2}$ for $\beta=2$, $c \geq 1$ for $\beta=1$ to guarantee a connected network. Accordingly, we indicate by $c'$ the prefactor corresponding to the dense regime. It follows that
\begin{eqnarray}
\frac{P_{n,1}}{P_{n,2}} \approx \left(\frac{2c}{n}\right)^{cn}\left(\frac{1}{2c'} - 1\right)^{c' n^{2}}\left(\frac{1}{1-2c'}\right)^{n^{2}/2} f_{n},
\end{eqnarray} 
where $f_{n}$ is a term accounting for all the intervening factorial terms. Noting that the third term in the right-hand side of the equation is always larger than 1, except when it is undefined for $c'=\frac{1}{2}$, we have that the overall scaling of the terms within parentheses is $e^{n^{2}}$. To calculate the scaling of the term $f_{n}$ let us apply the Stirling approximation $n! \sim \sqrt{2\pi n}n^{n}e^{-n}$ to each one of the four factorial terms, to find the overall scaling
\begin{eqnarray}
\frac{P_{n,1}}{P_{n,2}} \sim e^{n}.
\end{eqnarray} 
\rev{By normalizing out the microscopic timescale $\Delta t_{n,\beta}$, the dimensionless quantity $\tilde{T}_{n,\beta}=T_{n,\beta}/\Delta t_{n,\beta}$ represents the expected number of trials required to reach functionality, quantifying the combinatorial difficulty of the evolutionary search. Accordingly, we can use this result to calculate the ratio between the expected number of trials needed to sample functional modules in the two regimes as} 
\begin{eqnarray}
\frac{\tilde{T}_{n,1}}{\tilde{T}_{n,2}}\sim e^{-n}\frac{\epsilon_{n,2}}{\epsilon_{n,1}},
\end{eqnarray}
which provides us with a first interesting result: \rev{dense architectures are exponentially inaccessible in an evolutionary search unless counterbalanced by exponentially greater functionality, i.e., unless $\epsilon_{n,2} \sim \epsilon_{n,1} e^{n}$.}
\rev{Furthermore, since
\begin{eqnarray}
\frac{T_{n,1}}{T_{n,2}} \sim e^{-n}\frac{\epsilon_{n,2}}{\epsilon_{n,1}} \frac{\Delta t_{n,1}}{\Delta t_{n,2}},
\end{eqnarray}
the exponential barrier in $\tilde{T}$ carries over to $T$ unless $\Delta t_{n,2}$ is itself exponentially smaller than $\Delta t_{n,1}$, which is a highly implausible scenario. In realistic conditions, sparse architectures are not just advantageous, but almost inevitable, because the time to reach functional dense structures is overwhelmingly larger.}

The previous result is valid, under the same hypotheses, at each level of the hierarchy. To keep the discussion general enough, we can also assume that at each level $l$ of the hierarchy there are a distinctive probability $\epsilon_{n,\beta}^{(l)}$ and time unit $\Delta t^{(l)}$. Accordingly, the overall time needed to form a hierarchical modular structure like the one described by our simple model is given by
\begin{eqnarray}
T_{n,\beta}^{(\text{hier})} = \sum\limits_{l=1}^{\ell} T_{n,\beta}^{(l)} = \frac{1}{P_{n,\beta}} \sum\limits_{l=1}^{\ell} \frac{\Delta t^{(l)}}{\epsilon_{n,\beta}^{(l)}},
\end{eqnarray}
where $\ell=\log_{n} N$. What is the equivalent time needed to form the same structure of size $N$ by chance, by exploring the whole configurations space? Let us assume that the relevant parameters in this case are $\epsilon_{n,\beta}^{\star}$ and time unit $\Delta t^{\star}$, to find that
\begin{eqnarray}
T_{N,\beta}^{\star} = \frac{1}{P_{N,\beta}} \frac{\Delta t^{\star}}{\epsilon_{N,\beta}^{\star}}.
\end{eqnarray}
It follows that
\begin{eqnarray}
\frac{T_{n,\beta}^{(\text{hier})}}{T_{N,\beta}^{\star} } = \frac{P_{N,\beta}}{P_{n,\beta}} \frac{\sum\limits_{l=1}^{\ell} \Delta t^{(l)}/\epsilon_{n,\beta}^{(l)}}{ \Delta t^{\star}/\epsilon_{N,\beta}^{\star} },
\end{eqnarray}

Let us now focus on sparse networks ($\beta=1$), since they are clearly favored with respect to dense ones, with $N\gg n \gg 1$. \rev{It is possible to show that 
\begin{eqnarray}
\frac{P_{N,1}}{P_{n,1}} \sim e^{\frac{1}{2}(N^2-n^2)\log(1-p)} \sim e^{-N^2}.
\end{eqnarray}}

\rev{As a concrete illustration, consider the protein--protein interaction network in \textit{Bacillus licheniformis} WX--02, whose largest connected component includes $N=1,718$ proteins and $1,3057$ interactions, with average degree $\langle k\rangle\simeq 12.6$~\cite{han2016prediction}. Although the empirical network is reported to be scale--free rather than Erd\H{o}s--R\'enyi, this distinction is marginal for our argument: applying our estimate with $n=100$ and $p\simeq 0.009$ gives about $10^{-5.7\times 10^3}$, an astronomically small probability highlighting the combinatorial barrier against one--shot assembly. In more realistic, heterogeneous topologies the ratio may be somewhat larger, but still vanishing on any practical scale, reinforcing the advantage of hierarchical assembly.} To balance this factor it is necessary that $\Delta t^{\star}$ is astronomically small, a requirement that is extremely unlikely. We have to conclude that a hierarchical development of complex networks is favored and can help to overcome the problems arising from the unreasonably small likelihood of forming similar structures at once, by scratch. One could argue that other mechanisms could be considered, such as growth processes with preferential attachment and other mechanisms~\cite{barabasi1999emergence,krapivsky2000connectivity,krapivsky2001organization,caldarelli2002scale,krapivsky2005network}: the main obstacles to such models is that they assume that each node has global knowledge of the existing network at a given time~\cite{trusina2004hierarchy} or that they do not generate genuine hierarchical modular structures.

Note that the formation process considered here is simplified for sake of clarity. In a more realistic scenario, at any level $i=0,1,...,\ell$ the groups do not need to have the same size $n$ and such a size does not need to be the same across all levels. Adding stochasticity to this simplified model will add realism without necessarily improving the insights we have gained about the structural and temporal advantages of hierarchical modular networks.

Remarkably, such structures have also a functional advantage with respect to other topologies, facilitating the persistence of over time of a diverse set of pathways for information exchange~\cite{ghavasieh2020statistical,ghavasieh2024diversity}. In fact, by considering the ensemble of diffusive processes unfolding on the top of a network according to the diffusion equation
\begin{eqnarray}
\dot{x}_i(t) = \sum_{j=1}^{N} A_{ji} [x_{j}(t)-x_{i}(t)],
\end{eqnarray}
it is possible to define the density matrix~\cite{de2016spectral,ghavasieh2020statistical}
\begin{eqnarray}
\label{eq:rho_Z}
\boldsymbol{\rho}(t) = \frac{e^{-t \mathbf{L}}}{Z(t)}, \quad Z(t)=\text{Tr}\left[ e^{-t \mathbf{L}} \right],
\end{eqnarray}
where $\mathbf{L}$ is the combinatorial Laplacian ($L_{ij}= k_i \delta_{ij} - A_{ij}$), $e^{-t \mathbf{L}}$ is equivalent to a temporal evolution operator and $Z(t)$ is a normalizing factor playing the same role of a partition function~\cite{ghavasieh2023generalized}. The diversity of information pathways \rev{in response to external perturbations} is quantified by the entropy of the ensemble, which is formally equivalent to the von Neumann entropy $S(t)= -\text{Tr}\left[ \boldsymbol{\rho}(t) \log \boldsymbol{\rho}(t)\right]$, \rev{where $t$ plays a role similar to the inverse temperature $\beta=1/k_B T$ and can be also interpreted like the scale of fluctuations}. Note that, in agreement with classic and quantum statistical mechanics, \rev{the derivative $\chi(t) = -\dot{S}(t)$} provides a suitable generalized thermodynamic susceptibility~\cite{villegas2022laplacian}. Similarly, it is possible to introduce the generalized free energy $\mathcal{F}(t)=-\frac{1}{t}\log Z(t)$. In fact, the von Neumann entropy is maximum for a system consisting of disconnected nodes (e.g., a gas of non-interacting units) and decreases when the network tends to a fully connected graph, since the number of distinct information pathways reduces. Conversely, the generalized free energy $\mathcal{F}(t)$ is minimum for a system of disconnected nodes and increases when the network tends to a fully connected graph, since it is easier to transfer information in this setting~\cite{ghavasieh2024diversity}.

By considering the formation process of a network at time $t$ in terms of the difference between its entropy and free energy with respect to a baseline -- e.g., the state corresponding to a gas of non-interacting units -- it is possible to characterize the relative trade-off between the gain in transport properties ($\delta \mathcal{F}$) and the loss of diversity in \rev{response} ($\delta S/t$) by
\begin{eqnarray}
\eta(t) = 1 - \frac{\delta S(t)}{\delta \mathcal{F}(t)}.
\end{eqnarray}

\begin{figure*}[!ht]
\centering
\includegraphics[width=\textwidth]{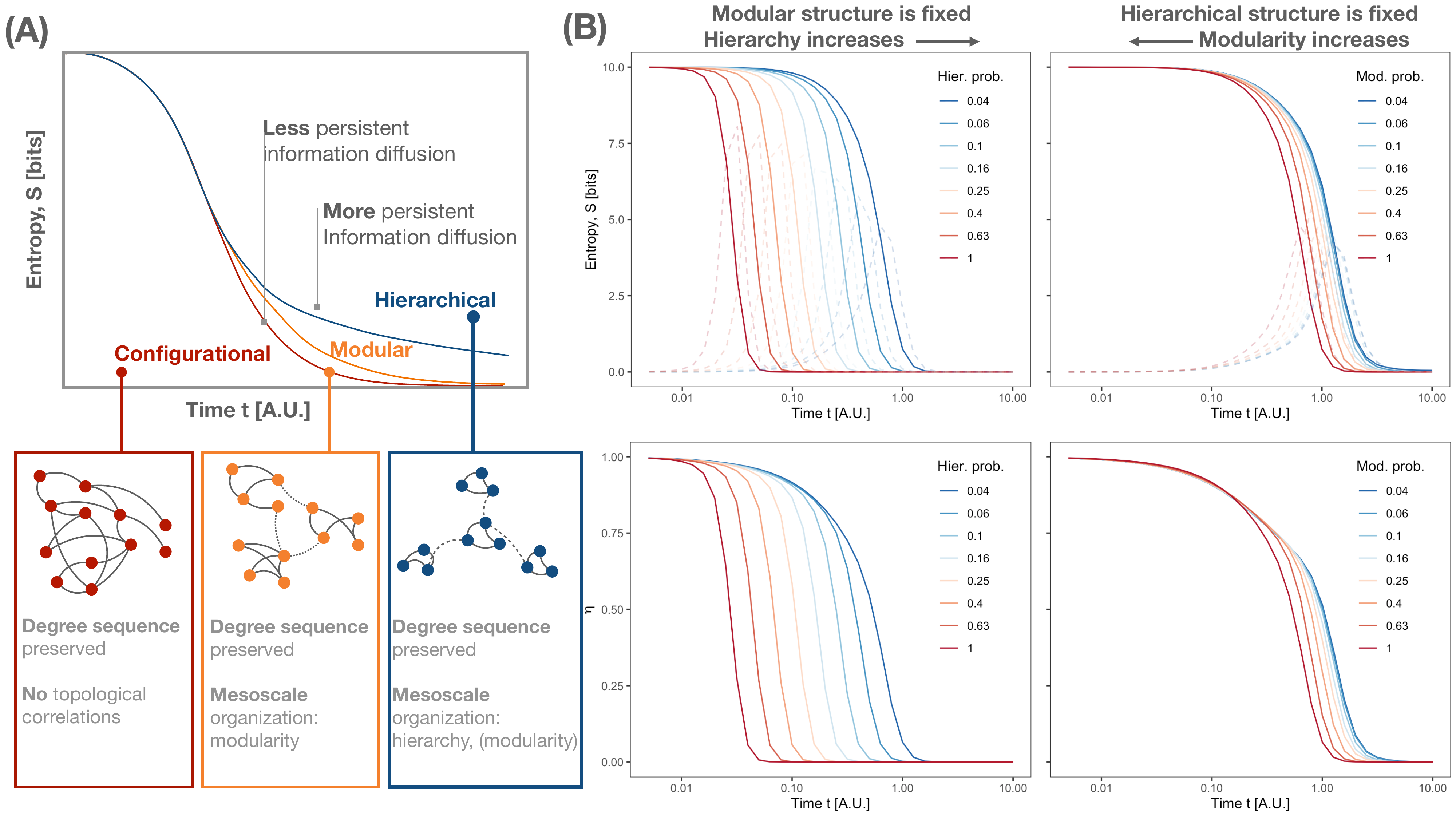}
\caption{\label{fig:synth_entropy}\textbf{Generalized thermodynamics of a synthetic benchmark}. (\textbf{A}) Sketch of the von Neumann entropy of complex networks as a function of diffusion time. The networks have the same average degree while being characterized by distinct connectivity patterns corresponding to distinct statistical ensembles. The configurational ensemble is not characterized by topological correlations, while the modular and hierarchical-modular ensembles provide different types of mesoscale organization. Hierarchical network ensembles have the highest information entropy~\cite{ghavasieh2020statistical}.
(\textbf{B}) The network model described in Fig.~\ref{fig:hierarchy} is used to generate synthetic benchmarks (size $N=1024$, levels $L=4$) with distinct hierarchical modular structure. Top: network von Neumann entropy versus time~\cite{de2016spectral} (solid lines) and corresponding generalized thermodynamic susceptibility (dashed lines) whose peak identifies the points of entropic phase transitions~\cite{villegas2022laplacian}. Bottom: network generalized efficiency~\cite{ghavasieh2024diversity}
Left: modular structure is fixed, while hierarchy is varied by tuning the probability ($p_{\text{hier}}$) to connect nodes on distinct levels; Right: hierarchy is fixed, while modularity is varied by tuning the ratio between the probability to connect within ($p_{\text{in}}$) and across ($p_{\text{out}}$) community.
}
\end{figure*}

Figure~\ref{fig:synth_entropy} shows how $S(t)$ and $\eta(t)$ change for the toy hierarchical model described in this section, highlighting how in hierarchical networks information exchange tends to be more persistent over long time scales, with respect to purely modular structures or to configurations without mesoscale and other topological correlations.

\section{The role of networks in major evolutionary transitions}\label{sec:major_evol_transitions}

While simple linear chains, regular lattices, trees or cliques can be considered as networks, broadly speaking, such topologies are barely useful to understand the structure and function of living systems. Instead, non-trivial topologies are usually more useful to make more accurate representations of biological systems, as we will see later. In fact, the non-trivial organization of biological units -- such as biomolecules, cells, tissues, organs, social and eusocial organisms -- endows the corresponding systems with specific features that are fundamental in facilitating new functions and structures essential for survival and adaptation. 

While evolution acts as a fundamental driver of change, complex networks likely serve as the primary agents of this change, facilitated by the dynamics of matter, energy, and information. Networks -- whether they are genetic, protein, neural, ecological, or social -- are the active components that translate evolutionary pressures into observable changes within and across organisms and ecosystems. Forcing an analogy with physics, we can see evolution as a fundamental law -- similar in spirit to the laws of thermodynamics -- and networks as the thermodynamic machines doing the work that can lead to complex adaptations and emergent behaviors~\cite{holland1992complex}. However, this is more than just an analogy since living systems, from molecular machines to ecological networks, can be literally understood as thermodynamic engines that convert energy from one form to another and perform work at the expenses of their environment~\cite{schrodinger2012life} (we will discuss more about this in Sec.~\ref{sec:model}).

In fact, biological networks are dynamic interactors in the evolutionary process: they are not static entities but are constantly changing in response to both internal and external pressures, a key characteristic of how living systems adapt and evolve. Networks mediate the effects of evolutionary forces by providing viable pathways within which genes, species, or individuals interact, thus shaping the path of evolutionary change and influencing which traits are likely to succeed or fail based on the network's current state and the environmental context~\cite{albert2003topology}. Accordingly, networks physically embody the connections and pathways through which biological functions are integrated and executed: they can be described as the structural implementors of evolution for selection through various adaptive processes. Finally, some network topologies can also be viewed as catalysts: they do not just passively support but actively accelerate evolutionary dynamics by facilitating the functional integration of sub-systems via rapid adaptations, feedback loops, and complex interactions that might be impossible in less interconnected systems or in systems that are overly interconnected. It is remarkable, in this regard, that mutualistic networks are regarded as the architecture of biodiversity~\cite{bascompte2007plant}, \rev{and how how perturbations to cross-feeding networks affect diversity~\cite{clegg2025cross}}.

The best way to appreciate the role of complex biological networks as agents of change is to analyze their role in the major evolutionary transitions (METs) as we know them~\cite{szathmary1995major,szathmary2015toward}. The METs are critical moments in the evolutionary history of the Earth, identifying new evolutionary levels where biological complexity increases significantly through the formation and evolution of novel forms of organizations, despite some limitations~\citep{omalley2016major}. 

We argue that such organizations can be naturally understood as network configurations that serve specific dynamic process and facilitate suitable functions. At this stage, the term “complexity” can be regarded as a proxy for stability, diversity, development of a variety of functions, nontrivial processes and behaviors that sustain and facilitate survival and adaptation.

In the context of METs, it might be beneficial to consider at least two classes of networks: the ones corresponding to the internal structures and processes of a single unit at a given scale (e.g., interaction and regulatory networks within a cell, a tissue, an organ, an organism, so forth and so on) and the ones corresponding to the external structures and processes involving those units. We are now ready to attack the following fundamental question: how can METs be conceptualized in terms of network theory? 

Any insightful answer to this question should (i) identify clear metrics for quantitative statements, and (ii) be able to demonstrate, in a principled way, how any divergence from the concept of complex network would inevitably lead to the disappearance of METs. Nevertheless, the opposite should not be neglected, i.e., how the disappearance of METs would inevitably lead to the disappearance of the intervening biological networks. In fact, the two concepts are tightly entangled and it is likely impossible to assess any causal priority to one direction or the other. This is compatible with the idea that is often summarized as “\emph{function from structure and structure from function}”, central in the context of adaptive networks. 

Since each new evolutionary level in METs corresponds to the spontaneous appearance or transformation of systems, it is essential to overview in detail the underlying phenomenology that requires networks for emerging function and organization. 
To this aim, we will use the 2015 classification proposed by E{\"o}rs Szathm\'ary~\cite{szathmary2015toward}, which builds upon the foundational framework developed with John Maynard Smith~\cite{szathmary1995major} and incorporates a broad corpus of influential studies (see references in~\cite{szathmary2015toward}, while we refer to~\cite{watson2016evolutionary} for a qualitative overview of evolutionary, ecological and developmental perspectives).

\subsection{Percolation phase transition}

\begin{figure}[!t]
\centering
\includegraphics[width=0.45\textwidth]{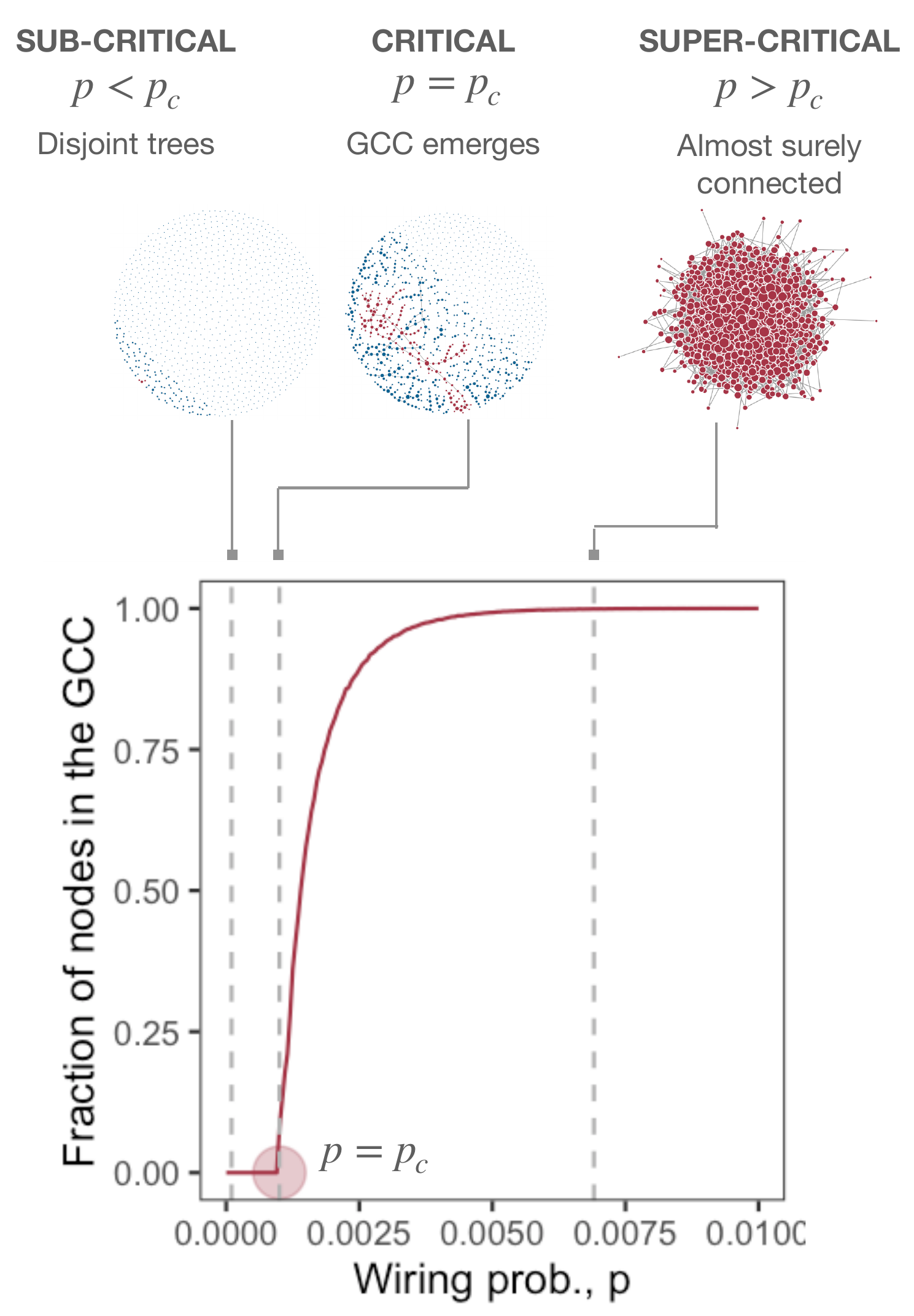}
\caption{\label{fig:percolation}\textbf{Structural percolation in complex networks}. Emergence of a giant connected component (GCC) as a function of the probability $p$ of connecting any pair of nodes in the Erd\H{o}s-R\'enyi model. The order parameter is the fraction of nodes belonging to the GCC, while $p$ acts as control parameter. The dashed lines correspond to three distinctive regions related to the percolation threshold $p_{\text{c}}$: (i) sub-critical, for $p < p_{\text{c}}$ the system is disconnected; critical, at $p=p_{\text{c}}$ the system undergoes a phase transition where a giant connected component appears; super-critical, for $p > p_{\text{c}}$ the system is connected. 100 independent random simulations of networks with size $N=1000$ have been performed.}
\end{figure}

A first fundamental argument in favor of the role of network description for all the processes where interactions and non-trivial organization is relevant, comes from the percolation transition that characterizes the emergence of a large connected component of interacting units. In a nutshell, not all interactions lead to a connected network and it has been widely studied under which condition they emerge. From the simplest random networks such as Erd\H{o}s-R\'enyi graphs~\cite{erdHos1960evolution} to more sophisticated structures such as Molloy-Reed random graphs~\cite{molloy1995critical} (see Sec.~\ref{sec:basic}), large connected components of units appear under specific structural conditions that guarantee enough connectivity. Regardless of the type of units involved, from molecules to whole organisms, the underlying organizational rule is the same: if a network is relevant to describe the behavior of a system, the conditions for its emergence from a set of randomly interacting agents reflect a percolation transition that, indeed, is related to the robustness of that system to targeted attacks and random failures of its constituents~\cite{artime2024robustness} (see Sec.~\ref{sec:architecture}). If we indicate with $\langle k^{n}\rangle$ the $n$--th moment of the degree distribution $P(k)$, where the degree $k_i$ simply counts the number of links adjacent to the $i$--th unit of a network, then the condition for the emergence of a giant connected component~\cite{molloy1995critical} is simply given by
\begin{eqnarray}
\kappa=\frac{\langle k^{2}\rangle}{\langle k\rangle}>2,
\end{eqnarray}
and it is related to the robustness of that network to randomly localized perturbations by
\begin{eqnarray}
\phi_{c}=1 - \frac{1}{\kappa - 1},
\end{eqnarray}
being $\phi_{c}$ the critical fraction of units to disrupt in order to disintegrate the giant connected component~\cite{cohen2000resilience}. Figure~\ref{fig:percolation} shows the percolation phase transition for the Erd\H{o}s-R\'enyi ensemble of complex networks, whose degree distribution is Poisson: accordingly, $\langle k^{2}\rangle =\langle k\rangle(\langle k\rangle+1)$ and the percolation threshold corresponds to $\langle k\rangle=1$ or, as equivalently shown in the figure, to $p_{\text{c}}=N^{-1}$.

However, as we are going to see in detail, connectedness coming from structural percolation is not the only feature required to characterize living systems: it is a minimum requirement. 

\subsection{Origins of life, the RNA world hypothesis and metabolism} 

Spontaneous formation of networks of autocatalytic molecules~\cite{vaidya2012spontaneous} provide a widely accepted foundational step in the transition from non-life to life~\cite{sole2025bifurcations}, where RNA not only acted as a genetic material but also catalyzed its own replication~\cite{gilbert1986origin}, a possibility experimentally reported by Bartel for ribozymes~\cite{bartel1993isolation}, possibly supporting a pre-protein biochemical framework in early life forms~\cite{noller1992unusual}. \rev{Long-term evolution experiments have shown that a single RNA is able to replicate by using a self-encoded RNA replicase, leading to a network of replicators~\cite{mizuuchi2022evolutionary}.}

In this context, the molecular networks~\cite{kauffman1969metabolic} essential to both genetic-first and metabolism-first hypotheses are crucial, with energy flow and cyclic transformations, as demonstrated by Morowitz~\cite{morowitz1966physical} (see also~\cite{schnakenberg1976network} and \cite{oster1973network} for an extensive review of how simple graph theory and generalized thermodynamics can be used for the analysis of non-equilibrium steady states in such systems). In the genetic-first scenario, these networks consist of nucleic acids that form self-replicating systems~\cite{orgel1992molecular} that can store and duplicate genetic information, endowed with catalytic functions. In the metabolism-first scenario, life originated from self-sustaining metabolic networks resonating with Morowitz's insights on external free energy sources as a driver of biological organization and with Kauffman's theory of autocatalytic sets~\cite{kauffman1971cellular}. Thus, whether focusing on genetic replication or metabolic cycles~\cite{watson2015understanding}, molecular networks are fundamental in delineating paths from chemical dynamics to the emergence of life~\cite{damer2020hot}. 

\rev{A further distinction is needed at this stage. Stoichiometric networks of small-molecule transformations constitute the biochemical transformation layer that sustains metabolism, whereas regulatory networks of macromolecules (DNA, RNA, proteins) provide a control layer superimposed on these transformations, enabling coordinated regulation and adaptive responses.}

\rev{Furthermore, it is important to note that such early networks were not necessarily hierarchical from the outset. Modeling work on genetic regulatory networks indicates that relatively flat, non-hierarchical architectures are more likely to arise initially, while hierarchical regulation emerges later as a derived outcome of increasing complexity and specialization~\cite{salazar2001phenotypicI,salazar2001phenotypicII,newman2000epigenetic}. These insights highlight that hierarchy should be understood as an evolutionary acquisition that enhances stability, robustness and fine-tuning, rather than as a primitive feature of such biological networks.}

Since these networks are heterogeneous and modular~\cite{jeong2000large,wagner2001small,von2002comparative,stelling2002metabolic,guimera2005functional,costanzo2010genetic}, it is plausible to wonder if and how this MET would be hindered by a dramatically structural change, e.g., into a dense random network or even a fully connected topology. Let us estimate the energetic, regulatory and maintenance costs in a reaction network with average degree $\langle k \rangle$ and $N \gg 1$ nodes, focusing on cycles of length $\ell$. Denote by $p(\ell)$ the probability that such cycles appear in random uncorrelated networks (see Sec.~\ref{sec:cycles} for mathematical details related to distinct network models). 

To estimate the energetic costs, consider the metabolic cost per reaction and the number of reactions per cycle. Assuming that each cycle of length $\ell$ involves $\ell$ reactions, and introducing a \rev{dimensionless scaling} exponent $\alpha$ to account for potential nonlinear scaling between metabolic costs and cycle length, the expected total energy cost across the network scales as
\begin{eqnarray}
C_{\text{total}}(\ell, E) \propto E \times \ell^{\alpha} \times p(\ell).
\end{eqnarray}
Here, $E$ represents the energy cost per reaction, and $\ell^{\alpha}$ captures complexities such as (non)cooperative interactions or (in)efficiencies in longer cycles.

Regulatory costs are related to the information processing required to manage and control these cycles within the network. If $b$ denotes the number of bits required per regulatory decision, and $j$ denotes the energy cost per bit (which has a theoretical minimum given by the Landauer's bound $k_{B} T \ln 2$ at temperature $T$~\cite{landauer1961irreversibility}) in joules/bit, the expected total regulatory energy cost for managing cycles of length $\ell$ scales as
\begin{eqnarray}
R_{\text{total}}(\ell, b, j) \propto j \times b \times \gamma(\ell) \times \ell^{\beta} \times p(\ell).
\end{eqnarray}
The \rev{dimensionless scaling} exponent $\beta$ accounts for the possibility that regulatory complexity does not increase linearly with cycle length\footnote{\rev{Note that this should not be confused with the exponent regulating the scaling between connectance and system's size in networks.}}, while the factor $\gamma(\ell)$ (with $0 < \gamma(\ell) \leq 1$) represents the efficiency gained through shared regulation, acknowledging that multiple reactions or cycles may be controlled by common regulatory mechanisms~\cite{bennett1982thermodynamics}.

Maintenance costs involve the energy required to repair, replace and maintain the enzymes and metabolites involved in the cycles. Introducing a \rev{dimensionless scaling} exponent $\delta$ to account for nonlinear scaling between maintenance costs and cycle length, the expected total maintenance cost scales as
\begin{eqnarray}
M_{\text{total}}(\ell, m) \propto m \times \ell^{\delta} \times p(\ell),
\end{eqnarray}
where $m$ denotes the average maintenance cost per reaction.

This estimation shows how network structure significantly influences energetic, regulatory, and maintenance costs: e.g., dense networks are characterized by an exponentially larger number of cycles than sparse networks (Sec~\ref{sec:cycles}), leading to higher costs. Similar differences are expected between random networks with homogeneous and heterogeneous degree distributions. The result does not strictly depend on the details of the chosen functional forms for the costs: the most important feature is their proportionality to the probability of appearance of cycles, which is plausible, and the positive scaling with $\ell$.

One might argue that maintaining cycles can be made more efficient by optimizing the interdependence of distinct cycles. Nevertheless, even under this assumption, substantial differences between sparse and dense networks, as well as between homogeneous and heterogeneous ones, persist. On the one hand, a large number of cycles is beneficial for the robustness of the underlying processes, as high density guarantees redundancy of pathways, making the system less fragile to random failures~\cite{wagner2005energy}. On the other hand, a large number of cycles is detrimental for energetic, regulatory, and maintenance costs: accordingly, maintaining dense homogeneous networks is energetically expensive. The trade-off between these competing effects\footnote{\rev{Here, competition refers to evolutionary selection acting on network architectures, e.g., favoring topologies that minimize energetic and regulatory costs while preserving function.}} calls for the selection of systems that achieve the minimum level of redundancy necessary to guarantee robustness with minimal expenditure for maintenance.

This balance aligns with the principles that every bit of information processed incurs an associated energy cost~\cite{bennett1979dissipation}, and that evolution favors network configurations optimizing energy efficiency while maintaining necessary robustness~\cite{wagner2005energy}. Consequently, biological systems tend to evolve modular and hierarchical network architectures that minimize costs while preserving functionality and adaptability.

\subsection{The Eukaryotic revolution} \label{sec:eukaryotic}

\rev{The origin of eukaryotic complexity, required for a new major evolutionary transition, has often been linked to symbiotic energetics, particularly the endosymbiotic association between proto-eukaryotes and mitochondria, which has been argued to provide the energetic foundation for larger and more complex cells~\cite{lane2010energetics}. Furthermore, increases in cellular complexity are typically accompanied by a reduction in the proteome fraction that can be allocated to core translational machinery, thereby imposing direct constraints on growth rate~\cite{hatton2019linking}. The acquisition of novel cellular features generally entails the evolution of ancillary infrastructure required for their assembly, quality control and maintenance, further reallocating resources away from reproduction and toward sustaining the expanded cellular architecture\footnote{\rev{In this context, nodes represent cellular compartments or molecular species, while edges denote exchanges of metabolites, genes, or signaling molecules across membranes. The emergence of new connections between previously independent units enabled division of labor, metabolic complementation and long-term cooperative stability in eukaryotic cells.}}~\cite{munoz2024energetic}.}

\rev{However, this is not the only possible pathway to increasing complexity: comparative analyses have shown that bacterial and eukaryotic cells overlap substantially in size and that no clear break is observed in allometric scaling of energy with cell volume~\cite{lynch2007frailty,lynch2015bioenergetic,schavemaker2025bioenergetics}. From this perspective, mitochondria may have accelerated the diversification of eukaryotes but were not strictly necessary for the initial transition.}

\rev{Equally, if not more, critical for the eukaryotic transition was the origin of the nucleus. The separation of transcription from translation allowed for unprecedented regulatory complexity, enabling splicing, RNA processing and sophisticated control of gene expression~\cite{lynch2007frailty}. This compartmentalization made possible the evolution of multilayered regulatory hierarchies, which are central to the coordination of increasingly complex cellular functions.}

\rev{Regardless of the specific pathway, the integration of unlike units such as mitochondria and host cells created a novel level of individuality based on complexity and interdependence~\cite{szathmary2015toward}, representing an egalitarian transition.}
 \rev{Hierarchical organization emerges through the coordination of pre-existing, functionally distinct systems (e.g., metabolic and genetic machinery of mitochondria and host cell) that must be integrated into a coherent whole~\cite{queller1997cooperators,szathmary2015toward,west2015major}. Unlike fraternal transitions (see Sec.~\ref{sec:multicellularity}), where units start as similar and later diversify, egalitarian transitions require compatibility and regulatory mechanisms to manage the interactions of heterogeneous partners. From a network perspective, such transitions require compatibility and cross-integration of distinct sub-systems, producing hierarchies “from the outside in” as coordination precedes differentiation, in order to form a functionally synergistic unit.}

The underlying topology was crucial for the development of compartmentalization and the nucleus, facilitating more complex genetic interactions. Accordingly, if compartmentalization is essential, not all random networks can achieve the same result: the type of heterogeneity required in this case cannot be limited to the degree distribution and calls for the appearance of topological correlations that lead to the formation of a non-trivial mesoscale organization~\cite{peixoto2019bayesian}. Alternatively, the probability that the homogeneous or heterogeneous random networks considered in Sec.~\ref{sec:cycles} produce modular structures by chance is negligible, suggesting that other mechanisms were at work and facilitated selection.

In fact, modularity allows for localized interactions that can be optimized for specific functions without disrupting other processes within the system. This specialization enhances both efficiency and effectiveness, allowing for complex functions to be managed within a controlled environment, at variance with a random non-modular system where this task is less likely to succeed. 

A functional module is, fundamentally, a distinct unit whose operational scope is clearly delineated from that of other modules within the system. This separation often arises from chemical isolation, achieved either through spatial localization or specific chemical interactions. For example, the ribosome -- which is a core module in protein synthesis -- encapsulates the biochemical reactions necessary for polypeptide formation within a single, spatially confined unit. Conversely, modules governing processes like chemotaxis in bacteria or mating in yeast leverage the chemical specificity of their components. These systems ensure functional isolation through precise interactions beginning with the selective binding of chemical signals -- such as chemoattractants or pheromones -- to their respective receptor proteins, followed by a series of highly specific interactions among signaling proteins. This specificity in chemical interactions underpins the modular architecture of these systems, allowing them to execute specialized functions effectively without interference, thereby enhancing the overall efficiency and adaptability of the organism~\cite{hartwell1999molecular}.

Remarkably, both compartmentalization and modularity facilitate evolutionary adaptations by allowing for changes or innovations to appear and be adopted in one part of the system without requiring global changes across the entire network, a crucial feature for evolution in complex environments~\cite{rives2003modular,caetano2019emergence}. Therefore, the fact that modularity in metabolic networks corresponds to different metabolic pathways~\cite{guimera2005functional} can reflect the compartmentalized functions within the cell that would be otherwise impossible with non-modular structures. Modularity creates localized functional units, enabling evolutionary flexibility and, together with hierarchical organization, reducing the systemic risks of failure or supporting asymptotically stable linear dynamics~\cite{variano2004networks}.

A related, and equally important factor, is that these modular systems are embedded within spatially structured environments. The spatial arrangement of these systems influences how evolutionary dynamics play out, further complicating the adaptive process~\cite{devadhasan2024competition}. Although spatially resolved datasets in molecular biology are transforming our understanding, they are still underutilized in evolutionary biology. Fortunately, mathematical models like evolutionary graph theory, which represent heterogeneous population structures, offer a way to link spatial complexity to evolutionary outcomes. Recent research demonstrates that even small extensions to multi-mutation fitness landscapes can lead to complex, emergent behaviors that cannot be captured by single-mutation models. These dynamics are shaped by the network's amplification and acceleration properties, which influence how populations navigate fitness landscapes: spatial segregation is a viable pathway to generate modules that later will integrate their function with the existing sub-systems~\cite{kuo2024evolutionary}. Modularity and spatial structure together provide a suitable framework in which evolutionary adaptations are both localized and context-dependent, driving the complexity and adaptability of living systems. 

Regardless if the underlying network is sparse or dense, it is possible to make an estimate of the possible number of modules that can be obtained by partitioning a network of $N$ units. This value is known as the Bell's number and it scales incredibly fast with system's size, asymptotically as $\exp(N(\log N - 1))$: for increasing $N$ this scaling allows for an uncountable diversity of possible non-overlapping modules and, accordingly, of possible functions. This huge space of possibilities provides evolution with a formidable basin of opportunities for exploring disruptive innovation to increase the complexity of living systems, providing the necessary “tools” to manage adaptation to very different environmental conditions.

In addition to the mechanisms described above, evolutionary processes may involve both modularization and de-modularization. Initially, discrete modules enable local exploration and adaptation, allowing novel functions to emerge with minimal disruption to the global network. Once these localized innovations are refined and prove advantageous, they are gradually integrated into the system: a de-modularization process that consolidates successful adaptations and promotes overall network robustness. This dual mechanism ensures that evolutionary innovations are both locally tested and globally optimized by selection, providing a flexible framework for managing the trade-offs between metabolism, regulation and maintenance in response to environmental changes.

\subsection{Multicellularity and the appearance of complex organisms}\label{sec:multicellularity}

The evolution of multicellularity requires the development of extensive intercellular communication networks, coordinating the behavior of individual cells to function as a cohesive unit, and of complex regulatory networks to activate or inhibit the expression of specific genes that endow cells with the required traits. \rev{This represents a fraternal transition, where initially similar units (cells of a clonal lineage) diversify through differentiation to produce novel hierarchies~\cite{queller1997cooperators,west2015major,ratcliff2012experimental,ratcliff2014experimental}. Here, hierarchical regulation arises “from the inside out” as developmental programs and gene regulatory networks progressively specialize cell functions and coordinate them at the organismal level. In contrast with the egalitarian case of eukaryogenesis (Sec.~\ref{sec:eukaryotic}), where distinct sub-systems are integrated, fraternal transitions rely on the emergence of new regulatory layers to manage related units, enabling division of labor and the stabilization of cooperative interactions.}

Shifting from traditional physics-based models to network-based approaches to grasp multicellular development, it is possible to map how cells exchange information. From gap junctions in animals and plasmodesmata in plants to fungi septum pores, such networks allow to understand cellular coordination across different kingdoms~\cite{jackson2017topological,jackson2017network}, unraveling the emergent circuitry required to sustain a multicellular organism. 
There are several processes that can sustain multicellularity, such as division of labor, where distinct cells specialize to functions that benefit the organism as a whole, or stress protection~\cite{jacobeen2018cellular}, where peripheral cells cooperate to shield internal ones from external stressors. Remarkably, it has been recently reported that sparse networks of cellular interactions in early multicellularity favors reproductive specialization and other functional specializations~\cite{yanni2020topological}.

\begin{figure*}[!ht]
\centering
\includegraphics[width=\textwidth]{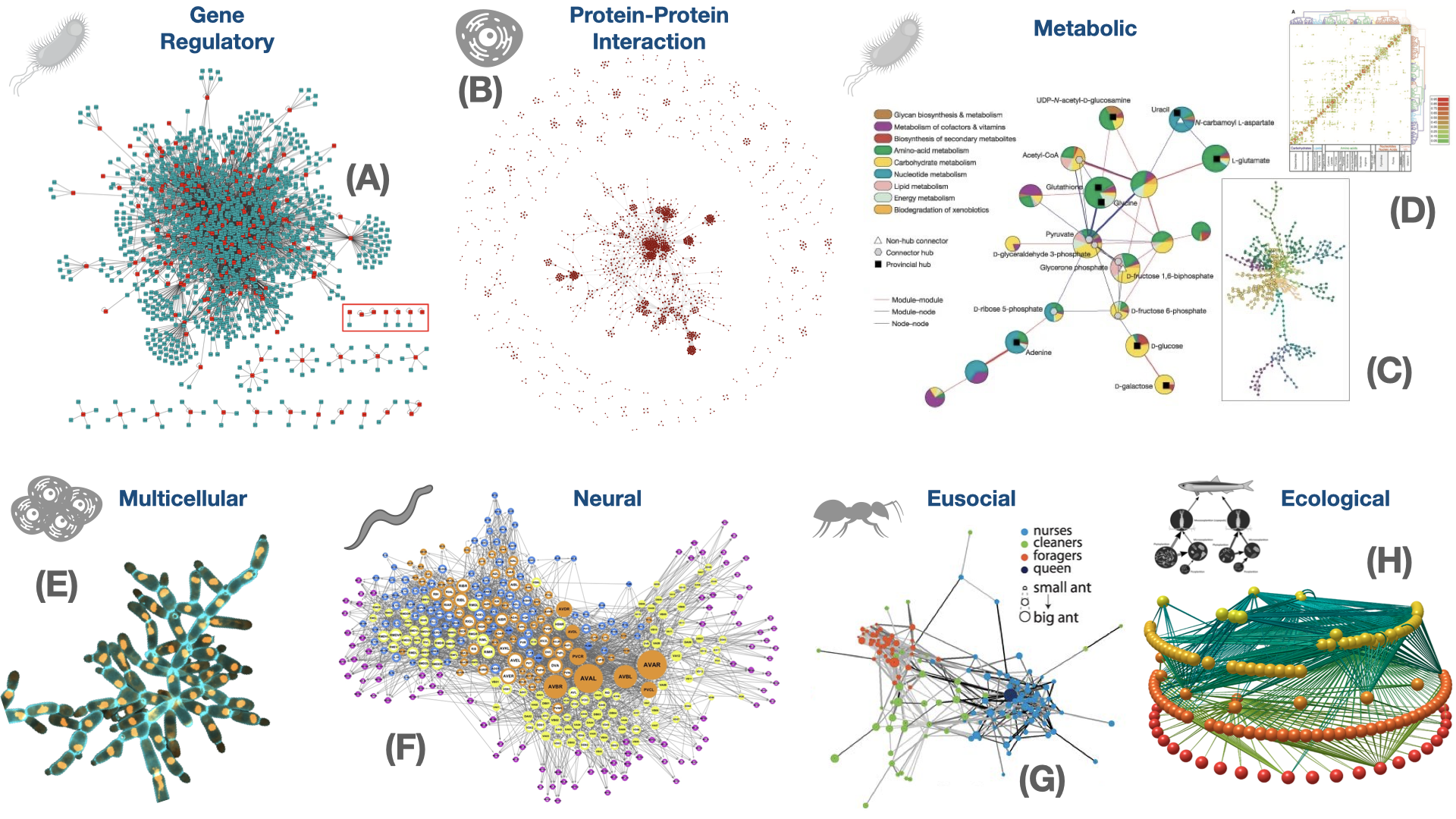}
\caption{\label{fig:empirical}\textbf{Emergent circuitry in living systems}. (A) Gene regulatory network of \emph{Escherichia coli}, where red colored nodes indicates transcription factor encoding genes; figure from~\cite{martin2016graphlet}. (B) The physical interactome of Saccharomyces cerevisiae; figure readapted from~\cite{collins2007toward}. (C) Metabolic network of \emph{Escherichia coli} and its functional organization into modules, which are represented as pie charts with colors indicating the involved metabolic pathways; figure from~\cite{guimera2005functional}. (D) Hierarchical organization of modules in \emph{Escherichia coli}; figure from~\cite{ravasz2002hierarchical}. (E) Multicellular yeast called snowflake yeast, obtained through many generations of directed evolution from unicellular yeast in the lab, captured using spinning disk confocal microscopy. Cells are connected to one another by their cell walls, shown in blue. Stained cytoplasm (green) and membranes (magenta) show that the individual cells remain separate. (F) Connectome of \emph{Caenorhabditis elegans}, where node size encodes neuron's degree and color codes cell type (sensory neuron, interneuron, motor neuron, muscle); figure from~\cite{yan2017network}. (G) Social network of an ant colony (no. 5, day 3), where the number of interactions and their average duration is encoded by edge width and darkness, respectively. Node color codes different ant roles, while node size is proportional to ant size. Figure from~\cite{mersch2013tracking}. (H) Cambrian food web (Burgess Shale), where nodes represent taxa, vertically organized by their trophic level, and links indicate feeding: figure from~\cite{dunne2008compilation}.
Panel (A) is licensed under CC-BY 4.0. Panels (D,G) reprinted with permission from AAAS. Panels (C,F) Reprinted with permission from Nature. Panel (E) Source: William Ratcliff, Georgia Institute of Technology; Credits: Anthony Burnetti, Ozan Bozdağ, and William Ratcliff, Georgia Institute of Technology. Panel (H) is licensed under CC-BY 4.0. Icon for panel (H) Source: Wikimedia; Credits: Zingone A, D'Alelio D, Mazzocchi MG, Montresor M, Sarno D, LTER-MC team (2019), licensed under the CC BY 4.0 Icons from Freepik.}
\end{figure*}

Such processes require, once again, the action of regulatory networks to be enabled, since transcriptional regulation is more flexible than the genetic component~\cite{lozada2006bacterial}. An emblematic example is provided by the recent experimental development of multicellular structures in yeast under controlled conditions. \emph{Saccharomyces cerevisiae}, typically a unicellular organism, can be induced through mutations to evolve into multicellular “snowflake yeast” structures (Fig.~\ref{fig:empirical}E). Under certain environmental conditions, the resulting organism demonstrates higher fitness than its singled-cell ancestor by effectively retaining nutrients when food is scarce~\cite{ratcliff2015origins}.

Multicellularity's capability to spawn a vast array of complex organisms adapted to diverse environments is greatly augmented by the differentiation of pluripotent cells into specialized types. This differentiation is sustained by complex gene regulatory networks that provide a sophisticated architecture that not only orchestrates cellular behavior but also creates a combinatorial explosion of potential phenotypic outcomes, allowing multicellular organisms to thrive under various selective pressures. Disturbances to such regulatory systems can deteriorate cellular cooperation, reversing multicellular functions to unicellular ones~\cite{chen2015reverse} or leading to the proliferation of cells that behave destructively~\cite{trigos2018evolution}, setting the stage for carcinogenesis.

Through optimal mechanical interaction networks, cells adapt their arrangement on elastic substrates to form tissues and function under varying physical conditions~\cite{noerr2023optimal}. Epithelial cells organize and maintain tissue integrity through a network of cellular interactions~\cite{escudero2011epithelial}. Cellular connectivity and network dynamics predict development, organization, and growth of multicellular clusters~\cite{yoshida2014genetic,nanda2024dynamic}. Overall, the available evidence highlights the role played by networks in shaping complexity and adaptability of multicellular organisms, providing a better understanding of how these life forms develop, function, and evolve under diverse conditions. Furthermore, it has been shown that multicellularity might be less unlikely than expected, with chemical networks regulating the complex oscillatory dynamics of intracellular processes in cell colonies that can naturally lead to multicellular organism~\cite{furusawa2002origin}, and it has been proposed that the genotype-phenotype mapping is a multilayer network~\cite{de2013mathematical,de2023more} connecting organismal complexity across scales~\cite{sole2024multicellularity}.

Remarkably, the evolutionary trajectory towards multicellularity is also intertwined to the development of circulatory systems, a solution addressing the challenge of efficient nutrient transport in organisms larger than unicellular ones. In fact, this adaptive innovation underscored the evolution of complex life forms by facilitating effective distribution systems that support expanded body sizes and intricate structures like organs. It has been recently shown that simple physical mechanisms can enable biophysical scaffolds alternative, and prior, to the ones which are genetically encoded~\cite{narayanasamy2024metabolically}.

Despite their significant metabolic and maintenance demands, the emergence of these circulatory systems summarizes a type of evolutionary innovation that, although energetically costly, provides a disruptive advantage crucial for survival in diverse environments. These systems reflect a meaningful integration of cellular signaling, genetic regulation and metabolic pathways, enhancing the organism's ability -- and efficiency -- to manage resources and respond to environmental disturbances more effectively.

These advanced systems are not isolated developments but are deeply integrated within the broader context of multicellular evolution, which is governed by the complex interplay of genetic, ecological, and thermodynamic forces. They illustrate how multicellular organisms not only overcome physical constraints through biological innovations but also leverage these systems to foster greater complexity and specialization, while enabling not just survival but also thriving in varied ecological niches and driving the formidable diversity of multicellular life forms observed nowadays~\cite{stoy2024adaptive}.

These circulatory systems can be viewed as dynamic transport networks crucial for coordinating multicellular activities and facilitating the sophisticated architecture required for complex multicellular life: the corresponding integrated systems are responsible for driving cellular cooperation, specialization, and information processing, definitely contributing to an organism's evolutionary success across spatial and temporal scales. Exemplary sub-networks of this type are the immune and the nervous systems, able to enhance cellular signalling capabilities to coordinate efficient responses to internal and external stimuli, as well as to process a vast amount of information which is critical for survival and adaptation. To this aim, the increased costs for sustaining and maintaining such system is well balanced by a largely increased robustness and resilience to environmental stressors.

\subsection{The evolution of social systems and eusociality} 

The growth of multicellular systems paved the way for the emergence of complex organisms. At this level of complexity, what type of innovation could further endow organisms with novel evolutionary advantages? The next significant leap is the formation of social groups, a phenomenon particularly exemplified by eusociality in insects~\cite{nowak2010evolution}, such as bees and ants, as well as the appearance of social relationships observed in larger animals, such as elephants, cetaceans, and certain primates. Remarkably, these complex social structures enable cooperative behaviors and social learning~\cite{cantor2013interplay,whiten2018pervasive} that significantly enhance survival, reproductive success, and resource utilization. Since the Cenozoic, such features represented a crucial evolutionary advancement over non-social animals~\cite{taniguchi2024sensory} in terms of information sharing via communication networks, collective defense mechanisms against predators~\cite{stanton2012early}, and foraging efficiency~\cite{cantor2013interplay}. 

Such structures influence, and are influenced by, a variety of mechanisms that range from chemical signaling to more sophisticated hierarchical and modular interactions, resulting in solutions such as eusociality~\cite{taniguchi2024sensory} -- i.e., cooperative behavior and division of labor -- while raising the necessity for shared decision-making~\cite{lusseau2007evidence} and conflict management to increase fitness, \rev{consistent with theoretical results showing threshold-dependent benefits of cooperation on spatial networks~\cite{fahimipour2022sharp}}. In fact, social networks predict task allocation, survival, activity patterns, and future behavior of honey bees~\cite{wild2021social}, as well as survival in bottlenose dolphins~\cite{stanton2012early} and \emph{Macaca sylvanus}~\cite{lehmann2016effects}. The rapid dissemination of crucial survival information, such as the location of food sources or the presence of predators, enhances survival strategies and resource utilization in bottlenose dolphins~\cite{lusseau2003emergent}. Network structure provides reproductive benefits, such as increased mating opportunities and more stable environments for raising offspring, enhancing juvenile survival rates~\cite{mann2012social}.

Overall, heterogeneous, modular, and hierarchical networks are often favored in social animal structures~\cite{grueter2020multilevel} due to their resilience to environmental stressors, efficiency in information communication, and adaptability~\cite{cantor2013interplay}. Modular networks consisting of densely interconnected clusters optimize information flow by allowing localized efficiency while maintaining overall communication~\cite{ramos2006complex,lusseau2006quantifying,sueur2011comparative,cantor2012disentangling,nandini2018group,balasubramaniam2018influence}. Hierarchical and multilevel organization enables role specialization, enhancing organizational efficiency in dynamic environments where adaptability is crucial~\cite{wittemyer2007hierarchical,wiszniewski2009social,wittemyer2009sociality,de2012comparison,nandini2018group}, minimizes redundant interactions and optimizes communication paths, reducing the energy and cognitive load required to maintain social connections and status in large, complex societies~\cite{mccomb2000unusually,dunbar2007evolution,wittemyer2007hierarchical,sallet2011social,bickart2011amygdala,shannon2013effects,noonan2014neural}.

Nevertheless, the emergence of a social structure imposes new trade-offs on social animals, despite its advantages. Novel ecological challenges and constraints~\cite{wittemyer2005socioecology,pocock2012robustness,suweis2013emergence,kefi2019advancing} are managed through the development of fission-fusion societies, complex adaptive systems~\cite{preiser2018social,madsen2024societies} where groups can flexibly merge or split based on resource availability, with stable subgroups forming higher-order structures, enabling adaptive responses to changing environmental and social conditions~\cite{lusseau2006quantifying,vance2009social,wittemyer2009sociality,de2011dynamics,archie2012elephant,nandini2017seasonal,nandini2018group}. \rev{In this context, collective intelligence can emerge as groups integrate and refine information across individuals and generations, facilitating cumulative improvements in problem-solving and decision-making efficiency~\cite{sasaki2017cumulative}.} This behavior notably maintains stable associations within constrained group sizes, balancing the costs and benefits of sociality~\cite{nandini2017seasonal}. For instance, altruistic behaviors necessary for collective benefit may come at the cost of individual fitness, potentially leading to internal conflicts or non-cooperative behavior where individuals benefit from the group without reciprocating~\cite{mccowan2008utility,lehmann2016effects}. The existence of relationships and interactions increases the vulnerability to disease transmission (see \cite{stroeymeyt2018social,weiss2021diversity} and references therein), the competition for limited resources within and across social groups~\cite{sueur2011comparative}, despite the benefits of cooperative foraging, and the maintenance costs of social bonds~\cite{de1988mechanisms,mccomb2000unusually,poole2005elephants,dunbar2010social,mccowan2011network,rekdahl2015non,tamarit2022beyond}. 

\subsection{Human societies and language} 

Complex networks play a fundamental role also in the development and functioning of human societies and their communication systems~\cite{newman2003social} (Fig.~\ref{fig:mechanism}). The evolution of human social structures has been significantly influenced by the advantages offered by hierarchical and modular networks, which optimize information flow, enhance cooperation, and facilitate the efficient transmission of cultural knowledge across generations~\cite{migliano2017characterization}. Human societies have harnessed the power of cooperation and complex communication, primarily through language, to adapt to ecological and social challenges in a coordinated way~\cite{szathmary2015toward}.

The emergence of hierarchical and modular social structures~\cite{grueter2020multilevel} is a feature observed in social animals and eusocial insects, and these structures similarly characterize human groups~\cite{peixoto2014hierarchical}. By efficiently organizing social relationships, they reduce the cognitive load on individuals, allow for division of labor and role specialization, and contribute to the resilience and flexibility of social groups. In contrast, simpler network structures like cliques, regular, or random graphs lack the adaptability and efficiency required for large-scale complex social systems~\cite{cantor2013interplay}, highlighting that social circuitry complements the logic of social systems.

However, while high interconnectedness within societies facilitates cooperation~\cite{marcoux2013network} and rapid information exchange, it also amplifies the risk of disease transmission~\cite{pastor2015epidemic} and can intensify competition for limited resources. This interconnectedness may lead to quicker depletion of shared resources and make societies more susceptible to widespread contagions, posing significant challenges that must be balanced against the benefits of social cooperation~\cite{murase2024computational}.

Complex networks specifically shape the formation and evolution of human communication, including non-verbal forms like visual communication. The complexity of human communication is managed through networks that organize linguistic elements into hierarchical and modular structures. This organization manages complexity by breaking down messages into manageable units: phonemes (basic units of sound) combine to form morphemes (smallest units of meaning), morphemes form words, and words form sentences to express complex ideas. These structures allow for both flexibility and stability in communication, making them adaptable to different contexts while maintaining coherence (see~\cite{koestler1967ghost} and Refs. therein). Modularity within these networks enables localized processing of specific linguistic elements while preserving global coherence, aligning with the brain's capacity for handling complex information~\cite{piantadosi2014zipf} and building intricate functional representations~\cite{huth2016natural}. In cultural transmission, these network structures play a crucial role in facilitating the efficient transfer of knowledge across generations, ensuring the preservation and evolution of cultural identity.

Not only do these structures impact communication, but they are also reflected in the architecture of the human brain~\cite{meunier2010modular,moretti2013griffiths,munn2024multiscale}. The brain has developed specific architectures to process language efficiently: specialized regions, particularly in the left-hemisphere frontal and temporal areas, form a dedicated core language network supporting language processing~\cite{fedorenko2024language,fedorenko2024language2}. Analysis of the functional connectome has shown that, within the human brain, energetic costs of signaling pathways in evolutionarily expanded regions are higher than costs in sensory-motor regions~\cite{castrillon2023energy}.

Overall, this neural wiring allows for efficient decoding and encoding of linguistic information, essential for rapid and context-dependent communication. One might wonder whether a more densely connected linguistic network -- where every word is linked with each other -- could function effectively: likely, such a fully interconnected linguistic network would be impractical due to overwhelming cognitive constraints. Studies have shown that human cognitive processing has limitations in handling vast amounts of information simultaneously~\cite{miller1956magical}. For instance, a language with 20,000 words that can be connected in all possible ways would result in about 400 million potential word pairs, far exceeding our processing capacity: it remains unclear whether grammatical rules and contextual understanding could enable humans to navigate such complexity.

Therefore, the hierarchical structure of language optimizes communication by conveying complex meanings through relatively simple combinations, offering an evolutionary advantage by potentially enhancing the evolvability of language, i.e., its capacity to adapt and develop increased complexity over time. This organization aligns with throughput and information capacity constraints in neural processing and reflects how the human brain specializes for this crucial function~\cite{fedorenko2014reworking}.

Building on this, the similarities in structural and functional organization of the human brain and language suggest the speculative, yet fascinating, idea that brain architecture and language networks co-evolved while balancing efficiency and flexibility trade-offs~\cite{deacon1998symbolic}. This co-evolution, which is still to be confirmed, would be far from trivial and not necessarily driven by optimization processes, as discussed by Simon~\cite{simon1955class} and Mandelbrot~\cite{mandelbrot1953informational}, who debated the emergence of complex statistical features like skew distributions and Zipf's law from stochastic processes and optimization. While Zipf's law describes an inverse relationship between word frequency and rank, reflecting the language network heterogeneity, word-word correlations and contextual dependencies are crucial in sentence construction, a complexity that simple stochastic models like random walks fail to capture~\cite{seoane2018morphospace}.

An intriguing, yet speculative, perspective emerges when we consider Conway's Law, which posits that organizations design systems mirroring their own communication structures. Extending this concept to human societies suggests that the structures of our language and even the wiring of our brains might reflect the social networks in which they develop. Just as organizations produce designs that echo their communication patterns, it could be hypothesized that human language and neural architectures could be shaped by patterns of social interaction. This idea, while requiring further empirical validation, reinforces the concept that complex network structures not only influence our communication systems but also potentially shape our cognitive processes, highlighting the profound interplay between social structures, language development, and brain architecture~\cite{conway1968committees}. A recent study further suggests that social interactions play a role in brain development among cooperative breeders, supporting the idea that social structures may influence cognitive and neural development~\cite{cerrito2024neurodevelopmental}. Another recent study highlights how neural activity across cingulate cortex and cerebellum is more correlated during social behavior than during non-social one~\cite{hur2024correlated}.

Overall, major evolutionary transitions cannot be just understood as changes in physical structures or new species arising, but in terms of changes in how biological and non-biological information is used to create new levels of biological organization. The emergence of circuitry that favor information exchange across scales cannot be disentangled from evolutionary dynamics, being a necessary step to increase complexity and specialization. 

In this section we have described the phenomenological perspective to highlight the observational features that should be reproduced by a suitable modeling framework. How can we model the broad spectrum of dynamics and behaviors characterizing the METs? In the following sections, we will first outline three fundamental features to characterize, in general, the architecture of living systems and we will formalize such insights into a multiscale model.

\section{The architecture of living systems}\label{sec:architecture}

To gain some insights about the architecture of living systems we have to introduce the concept of evolvability and, accordingly, we have to show how it is influenced by complex circuitry. Evolvability~\cite{dawkins1989evolution,kirschner1998evolvability,conrad1990geometry} is the capacity of a biological system to generate heritable genetic variation upon which natural selection can act, and it is fundamental to adaptability and long-term survival~\cite{wagner1996perspective,wagner2013robustness}, generating adaptive variations that enhance fitness in changing environments~\cite{pigliucci2008evolvability}. It should be noticed that it is a property of the genotype-phenotype map rather than a population one~\cite{hansen2006evolution} and, remarkably, it has been recently shown that evolvability can be experimentally evolved~\cite{barnett2025experimental} (see also~\cite{payne2019causes} for a review of experimental work).

Therefore, while adaptation refers to the process by which organisms improve their survival and reproduction in their current environment (e.g., through the natural selection of organisms with advantageous heritable traits), evolvability represents the ability of a population to generate those heritable phenotypic variations. Essentially, adaptation is the result of selection acting on the variation generated by an organism's evolvability.

Some key factors enhanced evolvability during major evolutionary transitions. For instance genetic modularity reduces pleiotropy by organizing traits into partially independent units, allowing traits to evolve independently without adversely affecting others~\cite{wagner1996perspective}. Regulatory networks influence adaptive capacity: upstream or highly connected genes can drive significant phenotypic changes~\cite{olson2012adaptive}, while redundancy of transcriptional regulation promote evolvability in regulatory networks such as ribosomal regulation in yeast~\cite{tanay2005conservation}. Developmental plasticity enables organisms to produce varied phenotypes in response to environmental stimuli, increasing the likelihood of finding adaptive solutions~\cite{pattee2019evolved}.

Overall, complex networks ranging from genetic regulatory networks and metabolic pathways to neural circuits play a critical role in promoting evolvability~\cite{jaeger2014bioattractors}. Their modular and hierarchical structures allow flexible adaptation by confining mutation effects to specific components, preventing widespread detrimental impacts~\cite{olson2012adaptive}. This architecture facilitates exploration of the adaptive landscape, enabling populations to navigate between different adaptive peaks and avoid evolutionary stagnation, a concept known as “extra-dimensional bypasses”~\cite{pigliucci2008evolvability}. Neutral spaces within these networks allow the accumulation of hidden genetic variation that can be revealed under changing conditions, providing an important reservoir for adaptive evolution~\cite{wagner1996perspective}.

Living systems can be envisioned as complex adaptive dynamical systems capable of processing information, adapting to changing environments, and exhibiting intricate interdependencies among their components. This perspective allows us to model biological entities across various scales -- from molecular to ecological~\cite{levin1998ecosystems} -- using dynamical systems theory, which describes their continuous and regulated responses to internal and external stimuli through variables like molecular concentrations, gene expression levels, or cellular activities evolving over time.

While the computational analogy provides valuable insights, it has also important limitations. Unlike deterministic computational systems, biological systems exhibit nonlinearity, stochasticity, and complex feedback loops: their actions are driven by biochemical processes optimized for survival and reproduction, not by solving mathematical problems. 

By accounting for both advantages and limitations of this framework, in the following we outline the three basic features that characterize the architecture of living systems and provide the physical, chemical and ecological ground for evolvability and to evolve evolvability, as discussed in the previous sections for the major evolutionary transitions.

\subsection{Interconnected structure: hierarchical modular organization} 

Complex systems such as living organisms are characterized by a network structure~\cite{boccaletti2006complex} which can be either emerging from their function or affect their function, the two possibilities not being mutually exclusive~\cite{berner2023adaptive}. The interconnectedness of biological networks also underpins the dynamical behavior of living systems, facilitating emergent properties such as robustness and adaptability. Networks such as protein-protein interactions, gene regulatory circuits, and metabolic pathways display behaviors that can be effectively understood by means of dynamical systems models. For instance, it has been shown that proteins within cells perform functions akin to information processing, exhibiting feedback loops, bistability, and other dynamic behaviors similar to engineered systems~\cite{bray1995protein}. Proteins act as computational elements by integrating multiple inputs and producing outputs based on allosteric modifications, functioning like logical gates such as AND or OR gates. This interconnected modularity enables adaptive and flexible responses to stimuli from a changing environment, out of equilibrium.

Similarly, metabolic networks further might exemplify this computational capacity: chemical reactions, such as the ones defining metabolic pathways, can perform computations, effectively processing biochemical inputs through enzymatic reactions to regulate metabolic fluxes~\cite{hjelmfelt1991chemical,hjelmfelt1992chemical}. Gene regulatory networks can also function like logical gates, integrating multiple inputs and making cellular decisions based on threshold mechanisms~\cite{shmulevich2002boolean}. 

All biological systems exhibit modularity, where distinct functional units independently process information, enhancing the system's ability to adapt without necessitating widespread structural changes~\cite{hartwell1999molecular}. \rev{Modularity creates redundant copies that can diverge and specialize -- e.g., facilitating duplication of genes and pathways --  enhancing evolvability and expanding the accessible functional repertoire of living systems~\cite{lynch2000evolutionary,teichmann2004gene}.} This modular organization, which is reflected in segregated--while-integrated structures is ubiquitous and often further organized into hierarchies~\cite{ravasz2002hierarchical,corominas2013origins}, allowing for localized optimization, facilitating evolutionary innovations and thus contributing to persistence, resilience and evolvability of biological systems~\cite{simon1961aggregation,nordbotten2018ecological}.

Network sparsity enables efficiency, and it is a critical aspect of these interconnected structures, that are much more likely than random expectation (see Sec.~\ref{sec:hierarchy}). Biological networks evolve to be sparse, minimizing unnecessary connections while maximizing functional efficiency, and balancing information exchange with response diversity \cite{ghavasieh2024diversity}. The emergence of topological features such as modularity, small-worldness, and heterogeneity aligns with maximizing the trade-off between information exchange and response diversity~\cite{ghavasieh2024diversity}, although further theoretical work is needed to confirm this insight.

\subsection{Adaptive behavior: redundant, plastic and ecological} 

Adaptability must be another hallmark of living systems, since it is closely linked to modularity, redundancy, and robustness. At variance with deterministic systems, living organisms must survive in variable environments and exhibit a certain level of adaptiveness to randomness and biological noise to keep operating and remain stable or functional, from cells~\cite{kitano2004biological,lozada2006bacterial,kitano2007towards,eldar2010functional} to ecosystems~\cite{pocock2012robustness,kefi2019advancing} and societies~\cite{fowler2010cooperative}.

Redundancy and degeneracy in biological systems further enhance adaptability in uncertain environments, by adding the ability of structurally different elements to perform the same function and contributing to increase robustness~\cite{ciliberti2007robustness} and flexibility~\cite{edelman2001degeneracy,tononi1999measures}. Robustness through redundancy ensures operational continuity despite individual component failures, and it can be achieved by structures with redundant pathways and heterogeneous connectivity able to maintaining functionality despite perturbations~\cite{albert2000error,cohen2000resilience,barkai1997robustness,pocock2012robustness,nordbotten2018ecological}. \rev{Yet, recent analyses indicate that this same redundancy can prolong relaxation toward equilibrium: when overlapping functional roles make the underlying interaction matrix ill-conditioned, the system's approach to steady state becomes computationally hard, leading to long-lived ecological transients~\cite{gilpin2025optimization}.}

Temporal plasticity, the ability to change network structures over time, also enhances efficiency and adaptability, with temporal organization being  fundamental to living systems and contributing to their dynamic adaptability~\cite{pittendrigh1961temporal,lozada2006bacterial}. Temporal networks can be controlled more efficiently and require less energy than static counterparts, offering advantages in terms of efficiency and adaptability~\cite{li2017fundamental}, and it has been shown that introducing temporal variation in network structure can lead to efficient synchronization even under resource constraints \cite{zhang2021designing}. Adaptive networks provide advantages for learning and optimization, where distributed strategies enable networked agents to interact locally, continually learning and adapting to changes, thus enhancing the system's adaptability~\cite{sayed2014adaptive,berner2023adaptive}.

\subsection{Interdependent structures and processes} 

The architecture of living systems is evolutionary and continuously shaped by interdependent structures and processes involving exchange and transformation of energy, matter, and information. Any biological system is rarely isolated from other systems~\cite{jacob1970logique}.
While the aforementioned interconnected structures are crucial, they must build interdependencies with each other and the biological processes they are involved to build the overall architecture~\cite{gao2012networks,de2023more}. At the cellular level, metabolic pathways and genetic regulatory circuits are coupled to manage cellular decisions through efficient processing of inputs. Functional modules -- spatially or chemically isolated units composed of several cellular components -- carry out discrete functions and are fundamental building blocks of cellular organization. These modules combine hierarchically into larger, less cohesive units, facilitating complex interdependencies. \rev{A remarkable organismal-level example comes from a recent study in mice showing that large-scale neural recordings reveal distributed representations of sensory, cognitive, and motor information across nearly all brain regions, rather than confined to localized cortical modules~\cite{international2025brain}, highlighting that function arises through distributed neural interactions rather than from  single centralized hubs~\cite{findling2025brain}.} This is also evident, for instance, in the neural-immune system interaction after myocardial infarction, where the brain recruits immune cells to modulate sleep, thereby limiting inflammation and promoting heart healing~\cite{huynh2024myocardial}.

On higher scales, energy and matter flow between components via complex networks~\cite{paine1980food}, \rev{where ecosystem functioning depends on community evenness and environmental context rather than species richness~\cite{maureaud2019biodiversity},} with information transfer coordinating processes to ensure adaptability and resilience~\cite{levin1998ecosystems,levin2008resilience}. Features like modularity and hierarchical organization characterize also these systems of systems and allow for localized optimization, facilitating evolutionary innovations, contributing evolutionary potential of biological systems~\cite{hartwell1999molecular} and conferring robustness and resilience by compartmentalizing disturbances~\cite{levin2008resilience}.

Therefore, every biological entity is a system of systems, each nested within higher-order systems and sometimes obeying rules that cannot be deduced solely by analyzing its individual components. This implies that each level of organization must be considered in the context of adjacent levels, as new characteristics emerge at every tier of integration, where concepts and techniques applicable at one level may not function at higher or lower levels~\cite{jacob1970logique}, \rev{consistent with models showing how micro-level co-evolutionary dynamics can generate emergent macroscopic adaptation~\cite{jensen2018tangled}.} This observation highlights the importance of hierarchical organization united by the logic of reproduction but distinguished by the means of communication and regulatory circuits inherent to each system~\cite{jacob1970logique}.

Remarkably, this perspective perfectly aligns with the concept of emergent phenomena~\cite{anderson1972more,artime2022origin} that arise between microscopic and macroscopic scales, the so-called “middle way”~\cite{laughlin2000middle}. Emergent properties are not predictable solely from understanding the individual components but result from complex interactions within the system: the presence of interdependencies across distinct subsystems~\cite{gao2012networks} and multiple contexts of interactions defines layers that can be effectively described by multilayer network models~\cite{de2023more}. 

Overall, interconnected structures facilitate the emergence of networks with complex topological features that favor evolvability, adaptive behaviors enable response and survival mechanisms, and interdependent structures and processes ensure the integration and functionality of biological (sub-)systems within and across scales.


\section{A unifying framework}\label{sec:model}

\subsection{Dynamical systems theory} 

By explicitly employing multilayer network models~\cite{de2013mathematical}, we can better represent the various interconnected structures and processes that characterize living systems. These models acknowledge that biological networks are not isolated but are part of a larger, interdependent framework~\cite{gao2012networks} where multiple types of interactions might occur simultaneously~\cite{de2023more}. This layered approach enhances our ability to analyze and predict system behavior, accounting for the complexity inherent in biological systems. In the following, let us refer to units as nodes of a network that defines a layer, which in turns characterizes a specific context for interactions. By coupling the behavior of multilayer systems with adaptive dynamics~\cite{berner2023adaptive} across different scales, we will present the backbone of a rather general mathematical model for the architecture of living systems which is able, in principle, to unify a variety of dynamics and behaviors (see, for instance, Refs.~\cite{deco2011emerging,adamski2020self,sole2024nonequilibrium,chen2024stability,breakspear2017dynamic,sole2024multicellularity,pastor2015epidemic,sole2024fundamental} and references therein). \rev{Note that, under structured or weak coupling, such multiscale dynamics can exhibit emergent time-scale separation, as shown in ecological systems~\cite{gilpin2025optimization}.}

At a given scale $\mathcal{S}$, given a set of $N_{\mathcal{S}}$ units and $L_{\mathcal{S}}$ layers, let us indicate by $x_{j\beta}(t; \mathcal{S})$ ($j=1,2,...,N_{\mathcal{S}}$, $\beta=1,2,...,L_{\mathcal{S}}$) the state of the system -- e.g., molecular concentrations, gene expression levels, species abundances, so forth and so on -- at time $t$. Furthermore, let us indicate with $u_{j\beta}(t)$ some external input signal or control applied to a biological system and with $\boldsymbol{\Theta}$ the set of parameters that dynamically change based on the system's states and/or external inputs. Finally, let $\boldsymbol{\Sigma}(t)$ indicate the state of the environment.

\begin{figure*}[!t]
\centering
\includegraphics[width=\textwidth]{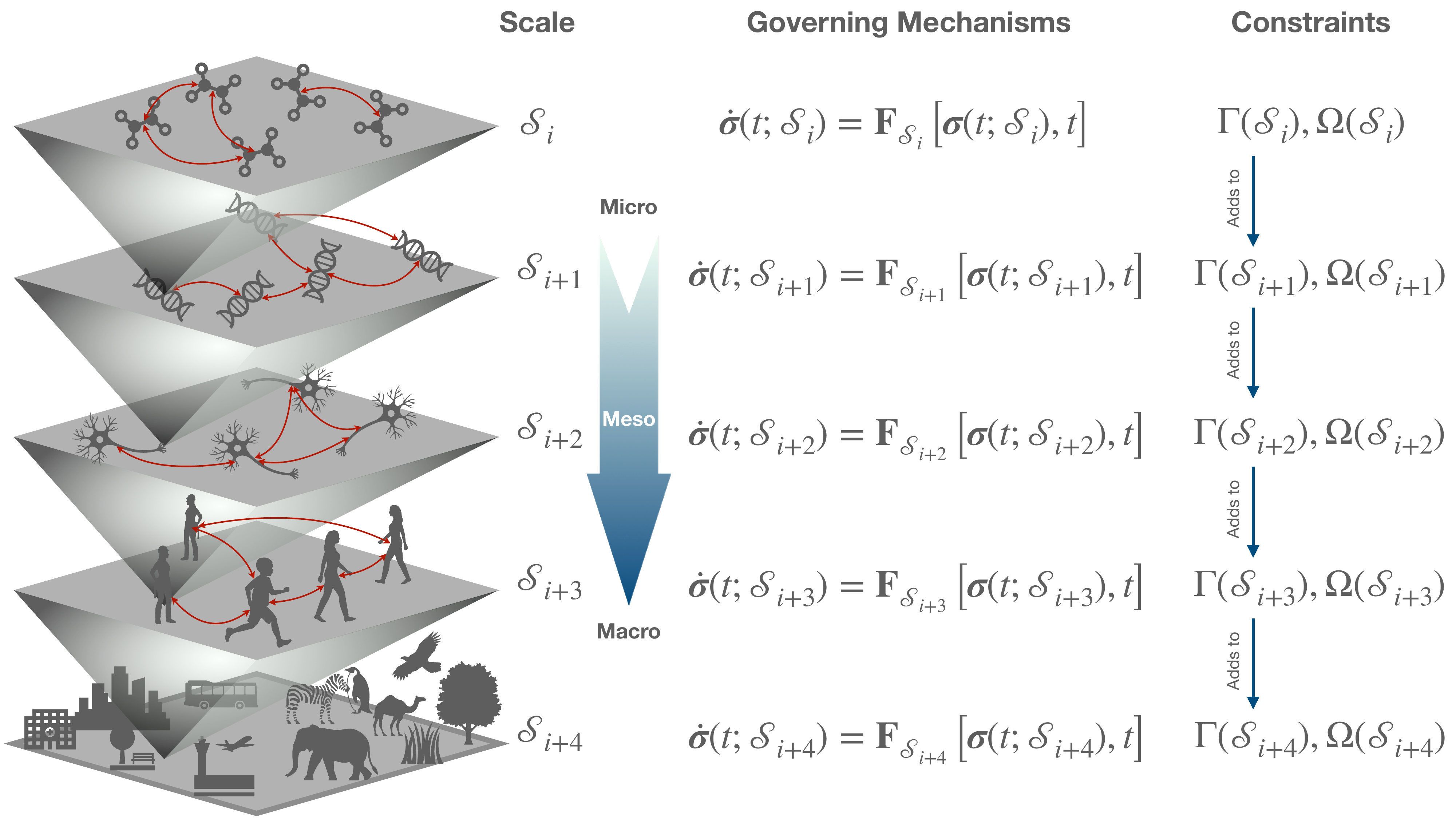}
\caption{\label{fig:scales} \textbf{Modeling dynamical systems across a hierarchy of scales}. As other complex systems, living systems are part of networks of networks which unfold across a hierarchy of spatial scales. At each scale $\mathcal{S}_{i}$ a specific set of novel constraints $\Omega(\mathcal{S}_{i})$ might appear and, together with constraints imposed by lower levels of organization ($\Gamma(\mathcal{S}_{i})$) and specific sets of non-autonomous and (potentially stochastic) differential equations, it is possible in principle to describe very complex dynamics of the involved units. Multilayer network modeling allows, in principle, to describe the behavior of such a complex multilevel system by accounting for interdependencies and influences within and across layers of organization. See the text for further details.}
\end{figure*}

A rather general model at that scale, not necessarily accounting for high-order interactions involving more than two units simultaneously but only pairwise interactions, can be formalized as

\begin{widetext}
\begin{eqnarray}
\label{eq:fast_dynamics_model}
\begin{aligned}
\frac{\partial x_{j \beta}(t)}{\partial t} & =\underbrace{f_{j \beta}\left(x_{j \beta}, t\right)}_{\text{Self-interactions}} + 
\underbrace{\sum_i \sum_\alpha g_{j \beta}\left[M_{j \beta}^{i \alpha}(t), x_{i \alpha}(t), x_{j \beta}(t), u_{j \beta}(t), \boldsymbol{\Theta}(t),\boldsymbol{\Sigma}(t), t\right]}_{\text{Pairwise interactions}}  \\
\frac{\partial M_{j \beta}^{i \alpha}(t)}{\partial t} & =\ell_{j \beta}^{i \alpha}\left[M_{j \beta}^{i \alpha}(t), x_{i \alpha}(t), x_{j \beta}(t), u_{j \beta}(t), \boldsymbol{\Theta}(t),\boldsymbol{\Sigma}(t), t\right] \\
\frac{\partial \boldsymbol{\Theta}(t)}{\partial t} & =\mathbf{h}\left[\boldsymbol{\Theta}(t),\boldsymbol{\Sigma}(t), \{u_{j \beta}(t)\}, t\right]\\
\frac{\partial \boldsymbol{\Sigma}(t)}{\partial t} & =\mathbf{s}\left[\boldsymbol{\Sigma}(t), \{x_{j \beta}(t)\}, t\right]\\
\end{aligned}
\end{eqnarray}
\end{widetext}
where we have omitted for notational simplicity the explicit dependence on the scale $\mathcal{S}$, and $M_{j \beta}^{i \alpha}(t)$ indicates the multilayer adjacency tensor~\cite{de2013mathematical} representing interactions between unit $i$ in layer $\alpha$ and unit $j$ in layer $\beta$ at time $t$: $M_{j \beta}^{i \alpha}(t)>0$ indicates the presence of an interaction or relationship, $M_{j \beta}^{i \alpha}(t)=0$ indicates its absence. In this model, each interaction is allowed to change over time as a function of the state of the system, the external input and the dynamical parameters, the latter being allowed to change as a function of the external input as well. Note that the system state depends on the state of the environment and, in general, the state of the environment is allowed to change with the state of the system, determining co-adaptive dynamics. By assuming that the physical space consists of interconnected spatial patches sustaining non-trivial dispersal patterns, spatial effects can be straightforwardly included in the model~\cite{levin1992problem,bascompte1995rethinking,moilanen2002simple,brechtel2018master,gross2020modern,nauta2024topological,sole2024nonequilibrium}.

Overall, the model is a non-autonomous system of coupled ODEs, which is able to account for the influence of the dynamical process on the structure and of the structure on the dynamical process, simultaneously, in response to the environment, external drivers (not necessarily environmental) and with possibly changing mechanisms (since the functions can vary over time). Note that this model can be made even more realistic by considering time-delayed interactions and effects, adequately included by means of delay parameters $\tau_{j\beta}^{i\alpha}$, or multi-body interactions. In fact, while one can assume that the dynamics of Eq.~(\ref{eq:fast_dynamics_model}) is Markovian in the state space, and governed by deterministic laws above the quantum scale (i.e., $\forall S_{i}$, with $i\gg 0$), the integration of a subset of variables may introduce both memory effects and noise sources, as shown by Zwanzig~\cite{zwanzig1961memory}. We will use this fact in Sec.~\ref{sec:unif_noneq_thermo} and Sec.~\ref{sec:unif_evo_dynamics}.

This modeling framework is rather complex, but at the same time it encompasses a broad spectrum of dynamics that are able to reproduce the variety of observed behaviors described in the previous sections. In the following, for sake of simplicity, let us introduce an overall state
\begin{eqnarray}
\boldsymbol{\sigma}(t;\mathcal{S})\equiv\left[ \mathbf{x}(t;\mathcal{S}), \mathbf{M}(t;\mathcal{S}), \boldsymbol{\Theta}(t;\mathcal{S}),\boldsymbol{\Sigma}(t;\mathcal{S}) \right]    
\end{eqnarray}
to describe the model in a more compact form as
\begin{eqnarray}
    \frac{\partial \boldsymbol{\sigma}(t;\mathcal{S})}{\partial t}=\mathbf{F}\left[\boldsymbol{\sigma}(t;\mathcal{S}),t\right].
\end{eqnarray}

\rev{Formally, this expression defines a non-autonomous vector field acting on the composite state space $\mathcal{X} \times \mathcal{M} \times \Theta \times \Sigma$, where each space describes, respectively, the variables $\mathbf{x}$, $\mathbf{M}$, $\boldsymbol{\Theta}$, and $\boldsymbol{\Sigma}$. The corresponding flow $\varphi_{t_0 \to t}:\boldsymbol{\sigma}(t_0;\mathcal{S}) \mapsto \boldsymbol{\sigma}(t;\mathcal{S})$ captures the system’s evolution in time.}

Note that we did not include, explicitly, a term to account for stochasticity: this extension can be surely made without changing the main message of this section, as well as time delays and higher-order interactions. At any scale $\mathcal{S}$, the model described is a non-autonomous system of coupled ODEs with set of constraints $\Gamma(\mathcal{S})$ specific to that scale: when changing the scale, new constraints may appear. This is a descriptive approach, but it raises the question: why do new constraints or laws appear at a new scale? These constraints cannot be deduced solely from lower scales, so why must they appear at all? 

In fact, physical constraints are fundamental at the lower levels, but then chemical laws emerge, adding to the physical laws. Subsequently, biological laws appear, possibly adding new constraints, and this continues up to social laws, which exist atop all preceding layers~\cite{jacob1970logique,anderson1972more,laughlin2000middle,artime2022origin}. Each time a new scale or level is introduced, there is a cumulative memory effect regarding constraints from lower scales: new units can appear at a new scale, but they cannot overcome the constraints imposed at the lower ones (e.g., the fundamental laws of physics remains valid even at the level of description of an ecosystem). Accordingly, given a hierarchy of scales $\mathcal{S}_{0}, \mathcal{S}_{1}, \dots, \mathcal{S}_{n}$, we have:

\begin{eqnarray}
\label{eq:hierarchy}
\Gamma(\mathcal{S}_{i})=\Omega(\mathcal{S}_{i}) \cup \Gamma(\mathcal{S}_{i-1}) , \qquad i=1,2,\dots,n
\end{eqnarray}

At each new scale $\mathcal{S}_{i}$, new units emerge, composed by aggregating units from lower scales, and these new units are subject to their specific constraints $\Omega(\mathcal{S}_{i})$ emerging at their scale $\mathcal{S}_{i}$, in addition to all the constraints from lower scales $\Gamma(\mathcal{S}_{j})$, $j=i-1,i-2,\dots$ (in simpler or different terms, a similar view is also present in \cite{polanyi1968life} and \cite{anderson1972more}). Here, $\Gamma(\mathcal{S}_{0})=\Omega(\mathcal{S}_{0})$ is the set of the fundamental physical laws of the Universe (Fig.~\ref{fig:scales}). 

Finally, we can include, at least in principles, interactions across distinct scales by introducing the overall state $\boldsymbol{\psi}(t)\equiv \left[ \boldsymbol{\sigma}(t;\mathcal{S}_1), ..., \boldsymbol{\sigma}(t;\mathcal{S}_{n}) \right]$ and describe its evolution as
\begin{eqnarray}
\label{eq:eta_dynamics}
    \frac{\partial \boldsymbol{\psi}(t)}{\partial t}=\mathbf{G}\left[\boldsymbol{\psi}(t),t\right],
\end{eqnarray}
which has a level of complexity that exceeds the ability of current methods for its analysis. Nevertheless, it provides a valid template to describe a wide variety of biological processes that characterize the multiscale architecture of living systems. 

The necessity for such a multiscale framework emerges for modeling phenomena characterized by feedbacks at a given scale, as in whole-cell models needed to accelerate model-driven biological discovery~\cite{karr2012whole}, and across scales. In the latter, at individual level it can enhance computational frameworks needed for precision medicine~\cite{de2025challenges}, while at ecological level it can model behaviors as the one recently reported to exist between social network structures and the composition of the human microbiome~\cite{beghini2024gut}. The study, which combines detailed social network mapping with microbiome sequencing in 1,787 adults across 18 isolated villages, reveals that microbial sharing is not confined to immediate family or household contacts but extends to second-degree social connections. Remarkably, individuals with higher social centrality exhibit microbiome profiles that converge toward the overall community signature, suggesting that the higher-scale social environment exerts a feedback influence on the microbial composition at the individual level, effectively modulating the coarse-grained state variables of the microbiome. Our framework, by integrating cumulative constraints across scales, is ideally suited to capture such feedback loops, illustrating how interactions at the social level can dynamically influence and be influenced by underlying biological processes.

At this point, one might wonder how it is possible that such an interconnected and interdependent system of systems can be resilient to perturbations, i.e., how emergent macroscopic behavior might be independent, to some extent, from the underlying microscopic details. For instance, to study the dynamics of microbial populations building the human microbiome it is possible to neglect the quantum mechanics of their constituent particles. For physical systems, this phenomenon is known as \emph{protectorate}~\cite{laughlin2000theory} and prevents minor changes at a lower scale to propagate dramatically at larger scales. The underlying idea is that certain emergent properties, such as superconductivity, fluid dynamics, or magnetism, can be described effectively by higher-level laws without needing to account for the complex, detailed interactions at the microscopic scale because the macroscopic description remains valid despite changes at the microscopic level. This is a form of effective theory where the higher-level behavior is separated from the lower-level behavior as long as certain conditions are met (e.g., that perturbations have no sufficient energy), and it apparently holds also for living systems~\cite{sole2024nonequilibrium}. It is still to be understood why the separation of scales applies for biological systems and the upper bound to the amount of changes that preserve this type of “biological protectorate”.

It is also worth remarking that an important limitation of the framework is due to the fact that traditional mathematical frameworks like ODE are inadequate for modeling emergent phenomena in biological systems due to their reliance on predefined variables and fixed phase spaces~\cite{kauffman2023third}. Accordingly, the continuous emergence of new adaptations (as new traits or new species) cannot be captured within these fixed-dimensional models. However, advanced approaches such as non-autonomous ODEs, infinite-dimensional dynamical systems, measure-valued differential equations and set-valued dynamical systems~\cite{hutson1992permanence} offer the flexibility to partially accommodate dynamic state spaces and emergent properties that cannot be readily captured by standard ODEs. 

While, on the one hand, these methods extend our modeling capacity, it should be acknowledged that important challenges persist in fully representing the unpredictability and open-ended evolution inherent in biological systems. On the other hand, it is not possible to shed light on the organizing principles of biological complexity without explicitly taking into account the role of fundamental constraints imposed by thermodynamics. 

In fact, according to the aforementioned Zwanzig's projection operator formalism~\cite{zwanzig1961memory}, when modeling systems at sufficiently fine-grained scales, all relevant degrees of freedom are explicitly represented. Upon integrating out fast variables (Eq.~(\ref{eq:fast_dynamics_model})), effective dynamics emerge in which thermodynamic behavior, memory effects, delays, so forth and so on, appear as phenomenological constraints. In this light, living systems can be understood as actualizing non-equilibrium thermodynamic constraints, in agreement with hierarchical organization (Eq.~\ref{eq:hierarchy}). These constraints, shaped by the multilevel structure of biological organization, define and delimit the landscape of possible system dynamics, making nonequilibrium thermodynamics a suitable framework for the analysis.

\subsection{Nonequilibrium thermodynamics} \label{sec:unif_noneq_thermo}

Biological systems are never in thermodynamic equilibrium, i.e., in a state where detailed balance holds and transitions between any two states occur at equal rates. Instead, living systems continuously exchange matter, energy and information with their environment -- for instance, through nutrient uptake or the absorption of light and heat -- which inherently disrupts this balance~\cite{nicolis1971fluctuations,prigogine1978time,fang2019nonequilibrium}.

These exchanges are essential for the emergence and maintenance of organized complexity in living systems, allowing them to function, evolve and maintain structure despite being subject to constant environmental perturbations~\cite{adamski2020self}, e.g., by means of self-organization~\cite{ashby1947principles,vonFoerster1992cybernetics,heylighen2001cybernetics}. Thermodynamics out of equilibrium and of irreversible processes~\cite{jarzynski1997nonequilibrium,seifert2012stochastic} provides a suitable framework to gain insights about fundamental constraints to self-organization of living systems, with early attempts started nearly half a century ago with Darwinian dynamics~\cite{bernstein1983darwinian,schneider1994life} for the emergence of macroscopic order.

In thermodynamic terms, we can describe the entropy of the system, $\mathcal{X}$, such as a living organism, at time $t$ and a given scale $\mathcal{S}$ as $S_{\mathcal{X}}(t;\mathcal{S})$. Similarly, the entropy of the surrounding environment, $\mathcal{E}$, at the same time and scale is given by $S_{\mathcal{E}}(t;\mathcal{S})$. The total entropy of the combined system and environment is then $S(t;\mathcal{S}) = S_{\mathcal{X}}(t;\mathcal{S}) + S_{\mathcal{E}}(t;\mathcal{S})$, and, according to the second law of thermodynamics, this total entropy cannot decrease over time~\cite{schrodinger1992life}: $\Delta S(t;\mathcal{S}) = S(t + \Delta t; \mathcal{S}) - S(t; \mathcal{S}) \geq 0$. This description is powerful, although an important drawback is that entropy is well established at equilibrium for a system with state variables. At equilibrium, entropy does not depend on the path taken to reach that state, whereas, far from equilibrium, it is no longer a state function. In fact, the dynamics of the a biological system includes the contributions of irreversible processes, such as metabolism, that produce entropy over time. The presence of gradients (e.g., temperature, chemical potential, so forth and so on) generates local variations in entropy, making difficult the definition of a global entropy if such variations are not negligible. One solution is to focus the attention on entropy production rates~\cite{nicolis1971fluctuations,prigogine1978time}, instead, although this has spurred further debate~\cite{anderson1987broken}, or to restate the second law of thermodynamics in terms of gradient destruction~\cite{schneider1994life} where dissipative structures -- and living systems -- can be understood as gradient dissipators. 

\onecolumngrid
\begin{mybox}[\textbf{Box B:} From Deterministic Dynamics to Nonequilibrium Thermodynamics]
\rev{
Equations~(\ref{eq:fast_dynamics_model})--(\ref{eq:eta_dynamics}) define a non-autonomous deterministic dynamics for the state $\boldsymbol{\sigma}(t;\mathcal{S})$, at scale $\mathcal{S}$. To describe coarse-grained mesoscopic states, we can introduce the slow observables $\boldsymbol{\psi}(t)$: according to the Mori-Zwanzig projection operator formalism~\cite{zwanzig1961memory}, the reduced dynamics is no longer deterministic and transforms into a non-Markovian stochastic integro-differential equation that captures the emerging dissipative and the fluctuating contributions  as
\begin{eqnarray}
\frac{d\boldsymbol{\psi}(t)}{dt} = \boldsymbol{\mathcal{A}} \boldsymbol{\psi}(t) + 
\int_{0}^{t} \mathbf{K}(t-s) \boldsymbol{\psi}(s)~ds + \boldsymbol{\xi}(t),
\end{eqnarray}
where $\boldsymbol{\mathcal{A}}$ is an effective drift operator accounting for systemic fluxes, $\mathbf{K}(t-s)$ a memory kernel, and $\boldsymbol{\xi}(t)$ a noise term (with $\langle \boldsymbol{\xi}(t)\rangle=0$) arising from integrating out unresolved degrees of freedom, which act as an effective thermal bath.

When the separation of time scales is strong, the memory kernel decays rapidly and its convolution can be approximated by an instantaneous friction term,
\begin{eqnarray}
\int_{0}^{t} \mathbf{K}(t-s)\boldsymbol{\psi}(s)ds \approx \boldsymbol{\Xi}\boldsymbol{\psi}(t),\qquad \boldsymbol{\Xi} = \int_{0}^{\infty} \mathbf{K}(s)ds.
\end{eqnarray}
If the eliminated variables remain close to equilibrium at temperature $T$, the fluctuation--dissipation relation gives $\langle \boldsymbol{\xi}(t)\boldsymbol{\xi}^{\intercal}(t') \rangle \approx 2k_B T(\boldsymbol{\Xi}+\boldsymbol{\Xi}^{\intercal})\delta(t-t')$, so that the reduced dynamics takes the Langevin form
\begin{eqnarray}
\frac{d\boldsymbol{\psi}(t)}{dt} = \left(\boldsymbol{\mathcal{A}} + \boldsymbol{\Xi}\right)\boldsymbol{\psi}(t) + \boldsymbol{\xi}(t),
\end{eqnarray}
Linearization near a steady state then recovers the Onsager linear relations, see Eqs.~(\ref{eq:onsager1})--(\ref{eq:onsager2}).

Finally, by coarse-graining the phase space into discrete regions, the mesoscopic dynamics then reduces to a Markovian process for the probability distribution 
$\mathbf{p}(t)=\{p_i(t)\}$ over effective states:
\begin{eqnarray}
\frac{d p_i(t)}{dt} = \sum_j \left[ W_{ij}(t)p_j(t) - W_{ji}(t)p_i(t) \right],
\label{eq:markov_approx}
\end{eqnarray}
where $p_i(t) \equiv \Pr\left[\boldsymbol{\psi}(t)\in \mathcal{C}_i\right]$ and $\mathcal{C}_i$ denotes the $i$-th coarse-grained region (or cell) of the reduced phase space of $\boldsymbol{\psi}(t)$. The antisymmetric probability currents
\begin{eqnarray}
\mathcal{J}_{ij}(t) = W_{ij}(t) p_j(t) - W_{ji}(t) p_i(t)
\end{eqnarray}
give the net fluxes between mesoscopic states, with an instantaneous entropy-production rate given by~\cite{schnakenberg1976network,seifert2012stochastic}
\begin{eqnarray}
\sigma(t) = \frac{1}{2}\sum_{i,j} \mathcal{J}_{ij}(t) \log \frac{W_{ij}(t) p_j(t)}{W_{ji}(t) p_i(t)} \ge 0,
\end{eqnarray}
which vanishes at detailed balance. Under local detailed balance, the stationary distribution $\boldsymbol{\pi}$ satisfies $W_{ij}\pi_j=W_{ji}\pi_i$, and the Kullback-Leibler divergence $\mathcal{D}(\mathbf{p}(t)\Vert\boldsymbol{\pi})$ acts as a Lyapunov function linking relaxation dynamics to entropy production, since $\dot{\mathcal{D}}(\mathbf{p}(t)\Vert\boldsymbol{\pi}) \le 0$.
}
\end{mybox}
\twocolumngrid

Within the general framework of non‐equilibrium thermodynamics, we can use the generalized state variable $\boldsymbol{\psi}(t)$ without assuming that it represents conventional thermodynamic quantities. Accordingly, we can hypothesize that there are $m$ state variables $\{\psi_1(t), \psi_2(t), \dots, \psi_m(t)\}$ that can serve as a coarse-grained description of the system's state, capturing effective degrees of freedom that describe deviations from a reference equilibrium state $\boldsymbol{\psi}^{\mathrm{eq}}$. By defining the deviation from such an equilibrium as
\begin{eqnarray}
\alpha_j(t) = \psi_j(t) - \psi_j^{\mathrm{eq}}, \quad j=1,2,\dots,m,
\end{eqnarray}
we have the basis for constructing effective thermodynamic forces as 
\begin{eqnarray}
\label{eq:onsager1}
X_j = \frac{\partial \Phi}{\partial \psi_j},
\end{eqnarray}
i.e., as the gradients of a suitable potential $\Phi(\boldsymbol{\psi})$, which can be interpreted as an effective free-energy-like functional whose gradients drive the evolution of the state. By introducing the conjugate fluxes $J_j$, and under the assumption that the deviations $\alpha_j(t)$ remain small (i.e., in the near-equilibrium or linear-response regime) and compatibly with the aforementioned protectorate, the fluxes are linearly related to the forces by
\begin{eqnarray}
\label{eq:onsager2}
J_i = \sum_{j=1}^{m} \mathbb{L}_{ij} X_j,
\end{eqnarray}
where the phenomenological coefficients $\mathbb{L}_{ij}$ satisfy Onsager's reciprocal relations $\mathbb{L}_{ij} = \mathbb{L}_{ji}$~\cite{onsager1931reciprocal1,onsager1931reciprocal2}, provided that time-reversal symmetry is maintained in the absence of external magnetic fields or similar symmetry-breaking influences, ensuring that higher-order nonlinear terms are negligible.

Consequently, the entropy production rate $\dot{S}(t)$ can be written as
\begin{eqnarray}
\label{eq:onsager3}
\dot{S}(t) = \sum_{j=1}^{m} J_j X_j \approx \sum_{i,j=1}^{m} \mathbb{L}_{ij} X_i X_j,
\end{eqnarray}
and the dynamics of the deviations $\alpha_j(t)$ can be described by
\begin{eqnarray}
\frac{\partial \alpha_j(t)}{\partial t} = \sum_{k=1}^{m} \mathbb{L}_{jk} \frac{\partial\Phi}{\partial \psi_k}.
\end{eqnarray}
Remarkably, this can be generalized to allow for memory effects and causal time behavior~\cite{zwanzig1961memory,zwanzig1972memory}, features which are also hallmark of living systems. \rev{See Box B for an overview of these concepts and how they are conceptually linked.}

\subsubsection{Correlations between biological systems and environment}

In the following, let $S(t)$ denote the total coarse‐grained entropy of the composite system at time $t$, \rev{evaluated under the assumption that the system can be considered in a locally equilibrated, quasi-stationary state, i.e., temporarily close enough to equilibrium for entropy to be well defined. The finite change $\Delta S(t) = S(t+\Delta t) - S(t)$ over a time interval $\Delta t$ then reflects the net effect of irreversible processes at the coarse-grained level, arising from the effective loss of microscopic information over $\Delta t$.}

\rev{However, biological systems are never isolated from their environments.} Their interaction with the environment often leads to correlations between the internal states of the system and the external states of the environment. 

\rev{Correlations reduce the overall number of accessible joint microstates, effectively constraining the joint probability distribution between system and environmental observables, so that the joint entropy $S(\mathcal{X},\mathcal{E})$ is smaller than the sum of marginal entropies, provided that they can be suitably defined at the coarse-grained level.} For uncorrelated systems, the total number of accessible microstates would be the product of the microstates accessible to the system and the environment. To account for these correlations, we introduce the correlation entropy $S_{\text{int}}(t)$ \rev{-- defined as the entropy reduction due to system-environment correlations, i.e. $S_{\text{int}}(t) = k_B \mathcal{I}(\mathcal{X}:\mathcal{E})\geq 0$ --}  leading to a modified expression for the total entropy
\begin{eqnarray} 
S(t) = S_{\mathcal{X}}(t) + S_{\mathcal{E}}(t) - S_{\text{int}}(t), 
\end{eqnarray}
\rev{which finds its energetic counterpart in the nonequilibrium free-energy balance $\Delta\mathcal{F} = W - k_B T\Delta \mathcal{I}$, derived in information thermodynamics~\cite{parrondo2015thermodynamics}.} Living systems that increase their organization over time -- at least within a limited time window -- lead to the condition $\Delta S_{\mathcal{X}}(t) = S_{\mathcal{X}}(t + \Delta t) - S_{\mathcal{X}}(t) \leq 0$. 

However, the decrease in the system's entropy must be balanced by an increase in the environment's entropy or in the correlation entropy, ensuring that the second law is not violated. The net change in the total entropy is then given by
\begin{eqnarray} 
\Delta S(t) = \Delta S_{\mathcal{X}}(t) + \Delta S_{\mathcal{E}}(t) - \Delta S_{\text{int}}(t) \geq 0, 
\end{eqnarray}
which implies
\begin{eqnarray} 
\Delta S_{\mathcal{E}}(t) - \Delta S_{\text{int}}(t) \geq -\Delta S_{\mathcal{X}}(t). 
\end{eqnarray}

Since $-\Delta S_{\mathcal{X}}(t) \geq 0$, this inequality shows that the reduction in entropy of a living system -- by consuming energy from the environment and doing work against it -- must be matched by the environment's entropy increase or by changes in the system-environment correlations. When the available energy is finite, the system can only maintain self-organization for a limited time. Eventually, when the energy is consumed, the system will no longer be able to sustain its organized structure. 

This brings us to the concept of free energy, which is intimately related to entropy and plays a central role in self-organization processes in living systems. While entropy governs the overall increase in disorder, free energy -- representing the usable energy available to do work -- helps explain how living systems maintain and increase their organization despite being in a nonequilibrium state.

\subsubsection{Free energy and information}\label{seq:free_energy}

The free energy of a system in thermal equilibrium is defined as $\mathcal{F} = E - TS$, where $E$ is the internal energy, $T$ is the temperature, and $S$ is the system's entropy. However, in nonequilibrium systems like biological organisms, the system is typically not in equilibrium with its environment, and the distribution of states deviates from the equilibrium distribution. Let $q(t) \equiv q(\boldsymbol{\psi};t)$ be the nonequilibrium distribution over the effective coarse‐grained state variables $\{\psi_1, \psi_2, \dots, \psi_m\}$, characterizing the system's microstate at time $t$. By indicating with $p\equiv p(\boldsymbol{\psi}^{\text{eq}})$ the equilibrium distribution, the deviation from equilibrium can be measured by the Kullback-Leibler (KL) divergence between $q(t)$ and $p$:
\begin{eqnarray} 
\mathcal{D}(q(t) || p) = \sum_i q_i(t) \log \frac{q_i(t)}{p_i}, 
\end{eqnarray}
where $q_i(t)$ is the probability of the system being in state $i$ at time $t$, and $p_i=e^{-E_i/k_B T}/Z$ is the corresponding equilibrium probability, with $Z=\sum_i e^{-E_i/k_B T}$ being the partition function. This KL divergence provides an information-theoretic notion of (pseudo-)distance between the nonequilibrium and equilibrium distributions, and can be related to the nonequilibrium free energy of the system as follows. It is convenient to define the nonequilibrium free energy as
\begin{eqnarray} 
\mathcal{F}_{\text{neq}}(q(t)) = \sum_i q_i(t) E_i + k_B T \sum_i q_i(t) \log q_i(t), 
\end{eqnarray}
which includes a generalized entropy term reflecting the actual distribution $q(t)$~\cite{kolchinsky2021work}. Since $\mathcal{F}_{\text{eq}} = -k_B T \log Z$ is the equilibrium free energy, the free energy difference between the nonequilibrium state and the equilibrium state is then related to the KL divergence as
\begin{eqnarray} 
\Delta \mathcal{F}(t) = \mathcal{F}_{\text{neq}}(q(t)) - \mathcal{F}_{\text{eq}} = k_B T \mathcal{D}(q(t) || p),
\end{eqnarray}
from which $\mathcal{D}(q(t) || p)=\Delta \mathcal{F}(t)/k_B T$, \rev{which has been shown to hold generally for any system with an energy function and a fixed bath temperature~\cite{kolchinsky2021work}}. Note that a similar result holds even when the system reaches a steady, nonequilibrium distribution, rather than one that evolves dynamically with time as $q(t)$. \rev{For systems maintained in nonequilibrium steady states (NESS) rather than canonical equilibrium, a closely related framework is provided by the Hatano-Sasa equality~\cite{hatano2001steady}, which generalizes the equilibrium relation: in this case, the steady‐state distribution $\pi$ defines a potential $\phi=-\log \pi$ and transitions between steady states obey an integral fluctuation relation linking the excess heat and changes in $\phi$.}

The relevance of free energy in the context of self-organization is clear: living systems must continuously consume free energy to maintain their organized structures. The process of reducing their internal entropy is possible only if they extract usable free energy from their environment, converting it into work that opposes the natural tendency toward disorder. As the KL divergence indicates how far the system is from equilibrium, it also quantifies the system's potential to do work and self-organize. A decrease in free energy signals the consumption of energy resources, ultimately leading to the exhaustion of the system's ability to maintain its organized state if external energy inputs are not sustained. Similarly, we can interpret the result from an information-theoretical perspective: by using information to keep the state far from equilibrium, the free energy difference is kept positive and self-organization is possible.

Therefore, understanding the dynamics of free energy in nonequilibrium systems is critical for modeling the self-organization and sustainability of biological systems. By tracking how free energy is consumed and how correlations evolve between the system and its environment, we can better understand the limits of organization in living systems and the thermodynamic constraints they operate under.

The link between the variation of free energy and information suggests that, to some extent, living systems might perform computational tasks (see Ref.~\cite{emmeche1994computational} and Refs. therein, as well as~\cite{bennett1982thermodynamics,bray1995protein,benenson2012biomolecular,czegel2022bayes,jaeger2024naturalizing}). While we will better describe this possibility in the Discussion, here we briefly focus on the specific task of learning, which is the most relevant to exploit available information to respond to perturbations and stimuli. Learning is often understood in terms of information processing, and can also be described as an evolution toward minimizing a loss function, where phenomena like replication and selection emerge as part of learning dynamics. Such dynamics align with principles of energy conservation, where the cost of learning is balanced against gains in adaptive fitness: in this context, adaptive systems exhibit behaviors that are analogous to the thermodynamic properties of renormalizable systems, and are able to learn through evolving, multi-scale adjustments~\cite{vanchurin2022toward}. Remarkably, this task is often thought to be achievable only via nonlinear processes. However, while nonlinearities often enhance computational capabilities, unlocking scalable and fault-tolerant learning via adaptive mechanisms~\cite{dillavou2024machine}, even linear dynamics can sustain learning by encoding information through parameters that influence physical processes~\cite{wanjura2024fully}. 

Learning, in the above terms, together with other features -- such as memory storage, information retrieval and even inference -- typical of computational systems is observed, to some extent, at any scales (Fig.~\ref{fig:scales}), \rev{since it is a form of adaptation}. However, the mechanisms through which learning is achieved must differentiate across scales, due to the distinct nature of involved systems and interactions. From short time scales, where individuals or ecosystems learn how to allocate resources or respond to environmental changes~\cite{lee2024constructing}, to evolutionary time scales, where genetic or epigenetic changes are acquired in response to environmental perturbations or stimuli~\cite{czegel2022bayes}, adaptation is driven by some form of learning process that is tied to natural selection.

\rev{These considerations invite comparison with formal approaches that cast evolution and adaptation themselves as inference or optimization processes. To this aim, a variational synthesis of evolutionary and developmental dynamics~\cite{friston2023variational}, extending the free energy principle (FEP)~\cite{friston2010free} to multiscale biological systems, has been proposed. Fast phenotypic processes and slow phylogenetic processes are treated as coupled stochastic dynamical systems whose paths minimize a free-energy-like functional. Fitness is defined as a path integral of phenotypic free energy and natural selection is interpreted as Bayesian model selection over generative models encoded by genotypes. The framework relies on the separation of temporal scales -- namely, the fast phenotypic (developmental/adaptive) and the slow phylogenetic (evolutionary) ones -- and, most importantly, on the construction of hierarchical Markov blankets\footnote{\rev{A Markov Blanket is the smallest set of variables enclosing all the information that can influence or be influenced by a system, such that, once conditioned on this set, the system becomes statistically independent from its environment.}} through renormalization operators that map between levels of description.

Our formalism, outlined in this section, shares with the FEP the use of information-theoretic measures -- specifically, the Kullback--Leibler divergence -- and the view that adaptation corresponds to the reduction of informational free energy. However, the two frameworks diverge in their ontological commitments and mathematical grounding. 

In the FEP, free energy quantifies the discrepancy between an organism's internal \emph{beliefs} and sensory evidence, assuming the existence of an internal generative model and well-defined Markov blankets separating system and environment. This inferential stance has been criticized for its limited falsifiability and for treating representational constructs as ontologically primary rather than emergent~\cite{bowers2012bayesian,bruineberg2022emperor,kirchhoff2018markov}. Recent enactive reinterpretations of active inference~\cite{ramstead2020tale} explicitly acknowledge this issue, proposing that generative and recognition models be understood not as internal representations but as formal descriptions of self-organizing, belief-guided dynamics. While this move alleviates some ontological tensions, it remains distinct from our perspective, which 
does not presuppose inferential mechanisms. Moreover, the term ``free energy'' in the FEP refers to a variational (information-theoretic) bound, which should not be conflated with thermodynamic energy or dissipative cost~\cite{buckley2017free,da2021bayesian}. 

By contrast, our approach is purely dynamical and thermodynamic: the same functional emerges as a measure of dissipative cost and constraint coupling among interacting components, without recourse to inferential or representational assumptions. Furthermore, whereas the FEP employs a formal renormalization-group construction to define hierarchical partitions, our coarse-graining procedure arises naturally from the explicit dynamics of fluxes, entropic forces and effective constraints across scales. While the FEP offers an elegant inferential framework unifying perception, learning and evolution, our unifying framework provides a physically grounded theory of multiscale organization and constraint emergence, from which observed inferential behaviour is interpreted as an emergent adaptive function resulting from underlying physical and dynamical constraints, rather than being presupposed as a universal organizing principle.
}

\rev{Finally, the framework developed here naturally raises the question of reciprocity between informational and physical levels of organization. Within this thermodynamic setting -- linking nonequilibrium free energy, information measures (e.g., $\mathcal{D}$), and self-organization -- empirical and theoretical studies indicate that biochemical and regulatory network architectures constrain the accessible modes of pattern formation by fixing reactive equilibria and flux balance. In mass-conserving reaction-diffusion systems, for instance, reactive nullclines and flux-balance subspaces provide a geometric scaffold for pattern selection~\cite{halatek2018rethinking,brauns2020phase,wurthner2022bridging}. A key open problem is whether the resulting spatiotemporal organization can, in turn, bias the evolutionary exploration of network space, i.e., whether persistent spatial or mechanical constraints act as filters that stabilize motifs compatible with specific nonequilibrium regimes. Establishing such bidirectional coupling empirically would extend the present thermodynamic framework toward a comprehensive account in which pattern-selection geometry and evolutionary dynamics co-determine the architecture of living systems.}

\subsection{Evolutionary dynamics} \label{sec:unif_evo_dynamics}

\subsubsection{Evolvability and robustness}

Evolutionary dynamics studies how populations change their traits over time under the influence of various forces such as natural selection, mutation and genetic drift. A central goal is to understand the long‐term behavior of traits -- what we term the “slow dynamics” -- which emerge after averaging over the rapid fluctuations occurring at smaller scales.

In this context, evolvability can be related to how sensitively the statistical properties of the system of systems change with a set of control parameters, such as $\boldsymbol{\Theta}(t)$ (see Eq.~(\ref{eq:fast_dynamics_model})). It is worth noting that even the underlying network structure can be considered as a control parameter to this aim: accordingly, in the following we will refer to $\boldsymbol{\Theta}$ as a general set which can perfectly match the one in Eq.~(\ref{eq:fast_dynamics_model}) or that is extended by accounting for structural and functional interactions.

In this unified framework, one can consider the probability distribution $q_{\Theta}(t)\equiv q(\boldsymbol{\psi}|\boldsymbol{\Theta})$ (see Sec.~\ref{seq:free_energy})), which characterizes the state of the system at time $t$ for a given control parameter set $\boldsymbol{\Theta}$ and define the Fisher information matrix as
\begin{eqnarray}
F_{ij}(\boldsymbol{\Theta}) = \mathbb{E}\left[ \frac{\partial \log q(\boldsymbol{\psi}|\boldsymbol{\Theta})}{\partial \Theta_i} \frac{\partial \log q(\boldsymbol{\psi}|\boldsymbol{\Theta})}{\partial \Theta_j} \right],
\end{eqnarray}
where the expectation is taken with respect to $q(\boldsymbol{\psi}|\boldsymbol{\Theta})$. This matrix measures the sensitivity of the probability distribution to infinitesimal changes in the parameters: accordingly, an operational -- yet simplified into a scalar -- measure of evolvability can be defined as
\begin{eqnarray}
\varepsilon(\boldsymbol{\Theta}) = \operatorname{Tr} \mathbf{F}(\boldsymbol{\Theta}),
\end{eqnarray}
where high values of $\varepsilon(\boldsymbol{\Theta})$ indicate that small variations in $\boldsymbol{\Theta}$ produce significant changes in the statistical behavior of $\boldsymbol{\psi}$, thus reflecting a high potential for evolvability.

It is worth remarking here that the above definition does not quantify robustness to perturbations $\delta \boldsymbol{\psi}$ in the state $\boldsymbol{\psi}$, which is typically assessed by the system's response. For instance, linear response along a trajectory in the state space is achieved through the analysis of the Jacobian matrix of Eq.~(\ref{eq:eta_dynamics}), namely
\begin{eqnarray}
\mathbf{J}(\boldsymbol{\psi}, t; \boldsymbol{\Theta}) = \frac{\partial \mathbf{G}(\boldsymbol{\psi}, t; \boldsymbol{\Theta})}{\partial \boldsymbol{\psi}}.
\end{eqnarray}
Conversely, $\varepsilon(\boldsymbol{\Theta})$ captures the sensitivity of the entire attractor landscape to changes in $\boldsymbol{\Theta}$, highlighting the capacity of the system to explore novel dynamical regimes in response to parameter variations, instead of state perturbations.

\subsubsection{Emergence of evolutionary dynamics}

Across different scales, or levels, we have considered a set of fast variables $\boldsymbol{\psi}(t)$ described by Eq.~(\ref{eq:eta_dynamics}), characterizing the rapid fluctuations in a biological system of systems. These fast dynamics may represent, for instance, molecular or cellular processes that evolve on time scales much shorter than those relevant for evolutionary change. Therefore, by coarse‐graining these fast degrees of freedom, one obtains effective dynamics for macroscopic -- i.e., evolutionary -- variables, such as population-averaged traits: in practice, the fast fluctuations determine an effective environment in which the slow variables evolve.

The choice of slow variables is usually guided by both empirical accessibility and theoretical considerations, where quantities such as the average trait value, denoted by $\langle \mathcal{A} \rangle$, are natural candidates because they capture the aggregate behavior of the population. For instance, even evolvability -- from classical ones~\cite{houle1992comparing,hansen2006evolution,jones2007mutation} to $\varepsilon(\boldsymbol{\Theta})$ -- can be seen as en evolvable trait that emerges from the fast dynamics.

It is worth remarking that the rate of change of these slow variables is constrained by thermodynamic and information-theoretic principles. For instance, the Fisher information quantifies the intrinsic speed at which a probability distribution evolves and imposes an upper bound on the rate of change of any observable. In nonequilibrium thermodynamics, uncertainty relations set limits on entropy production and energy dissipation~\cite{nicholson2020time}, thereby constraining the evolutionary speed. 

A key mathematical representation in evolutionary dynamics is the Price equation~\cite{price1970selection,price1972extension,price1995nature,frank1997price,frank2012natural}, which in continuous-time can be written as
\begin{eqnarray}
\frac{d\langle \mathcal{A}\rangle}{dt} = \langle \dot{\mathcal{A}} \rangle + \operatorname{cov}(\mathcal{A}, \phi),
\end{eqnarray}
where $\phi$ represents the effective growth rate or fitness. 
The first term, $\langle \dot{\mathcal{A}} \rangle$, accounts for any explicit time dependence of the trait $\mathcal{A}$, while the covariance term, $\operatorname{cov}(\mathcal{A}, \phi)$, captures the effect of differential reproduction (i.e., natural selection) on the average trait. 

\rev{It is important to stress, however, that the Price equation is a statistical identity that partitions the rate of change in the mean trait into covariance and expectation terms, rather than a dynamical law of evolution in itself~\cite{price1972extension,van2005use}. The equation is not specific to natural selection and can describe any process of change in a population. Accordingly, except for special cases, the Price equation is dynamically insufficient: it does not, on its own, generate evolutionary trajectories or specify the causal mechanisms of change~\cite{gardner2008price,frank2012natural}. To obtain predictive dynamics, one must add biological assumptions about replication, inheritance and variation, as is done in the replicator-mutator formulation, whose continuous-time version is also known as the Crow-Kimura equation~\cite{kimura1964diffusion,crow1970introduction},} when one includes additional terms to account for mutation~\cite{page2002unifying,traulsen2006coevolutionary,wakano2017derivation}:
\begin{eqnarray}
\label{eq:crow-kimura}\dot{x}_{i}(t) &=& \sum_{j=1}^{N_{\text{types}}} x_{j}(t) Q_{ji} \phi_{j}(t) - x_{i}(t)\Phi(t),\\ 
\Phi(t) &=& \sum_{i=1}^{N_{\text{types}}} x_{i}(t) \phi_{i}(t),
\end{eqnarray}
where $x_{i}(t)$ denotes the frequency of occurrence of trait $i$ in the whole population ($\sum_i x_i(t)=1$), $\phi_i(\{x_{k}\},t)$ is the fitness for that trait, $Q_{ij}\geq 0$ is a transition matrix (i.e., $\sum_{j}Q_{ij}=1$) encoding how trait $j$ is produced by type $i$ (e.g., because of random mutations), and $N_{\text{types}}$ is the total number of types considered. The replicator-mutator dynamics describe how both selection (through the covariance) and mutation (via additional transition probabilities) drive the evolution of the trait distribution. Evolvability, in this context, quantifies the potential of a population to undergo evolutionary change and it is directly related to the variance in fitness and to the constraints imposed by the underlying thermodynamic speed limits. High variability in fitness can serve as a resource, enabling rapid evolution, but only within the bounds dictated by non-equilibrium fluctuations and information-theoretic limits imposed by thermodynamics to quantum and classic systems~\cite{nicholson2020time,garcia2022unifying}, which have been recently shown to also apply to  biological ones~\cite{adachi2022universal,hoshino2023geometric,garcia2024limits}. 

By integrating out fast-scale fluctuations and employing both thermodynamic and information-theoretic constraints, we can arrive at a self-consistent picture of evolutionary dynamics. Accordingly, the replicator-mutator equation naturally emerges and sets the stage for understanding evolvability in terms of fundamental rate limits. 

It is natural to wonder why this equation is so relevant and ubiquitous to describe evolutionary dynamics in a variety of contexts, from population genetics to autocatalytic reaction networks, language learning, complex social behavior and economics~\cite{komarova2004replicator,safarzynska2010evolutionary,venkateswaran2019evolutionary}, and if it can be derived from some constrained variational approach, as other fundamental laws, without assuming \emph{a priori} its form.

To derive the replicator-mutator equation from first principles, we can use the maximum caliber approach~\cite{jaynes1980minimum,otten2010maximum,presse2013principles,ghosh2020maximum}. Max-caliber extends the principle of maximum entropy~\cite{jaynes1957information1,jaynes1957information2} to trajectories, providing a suitable variational framework to infer dynamical laws, far from equilibrium, from limited information and under the biological constraints on observable fluxes, such as average growth rates -- i.e., the fitness fields $\phi_{i}$ -- and mutation rates, i.e., the matrix $Q_{ji}$. 

To this aim, let us define the state of the system on two different levels. At the microscopic level, each individual in the population is assigned a type (or trait) and can transition between types over time. Following the approach proposed by Filyukov and Karpov~\cite{filyukov1967description,filyukov1967method} we denote a microscopic trajectory consisting of discrete states by $\Gamma = \{ i_0, i_1, \dots, i_T \}$, where $i_t$ represents the type of an individual $i$ at time $t$, and assume that the dynamics in the state space is Markovian:
\begin{eqnarray}
\mathcal{P}(\Gamma) = \pi_{i_0} P_{i_0,i_1}P_{i_1,i_2}\dots P_{i_{T-1},i_T},
\end{eqnarray}
where $P_{j_{t-1},j_t}$ is the elementary transition probability, $\mathcal{P}(\Gamma)$ indicates the probability distribution over the ensemble of trajectories with temporal length $T$ and $\pi_{i_0}$ is the stationary probability that the initial state $i_0$ is occupied (note that $\sum_i \pi_i = 1$). To simplify the notation, let $P_{ij}(t)$ indicate the probability that an individual currently in type $j$ at time $t$ transitions to type $i$ at time $t+1$, noting that it captures the behavior at the level of individual transitions. For a given microscopic path, the number of such transitions from state $j$ to state $i$, already happened at time $t$, is given by 
\begin{equation}
N_{\Gamma}(j\to i,t) = \sum_{t'=0}^{t-1} \delta_{j,j_{t'}}\delta_{i,j_{t'+1}},\nonumber
\end{equation}
where $\delta$ is the Kronecker delta. The Max-Caliber approach consists of maximizing the path entropy
\begin{equation}
S = -\sum_{\Gamma} \mathcal{P}(\Gamma) \log \mathcal{P}(\Gamma),
\end{equation}
subject to the following constraints: (i) the normalization $\sum_{\Gamma} \mathcal{P}(\Gamma) = 1$, and (ii) the biological mechanisms encoding the mean flux, which can be expressed as
\begin{equation}
\langle N(j\to i,t) \rangle = \sum_{\Gamma} \mathcal{P}(\Gamma) N_{\Gamma}(j\to i,t) \equiv \pi_j(t)W_{ji}(t),\nonumber
\end{equation}
where the elementary transition rate $W_{ji}(t)=\frac{Q_{ji}\phi_j(t)}{\Phi(t)}$ introduces the intervening biological mechanisms, namely the proportionality with the fitness of type $j$ and the probability kernel for mutation (with $\sum_i Q_{ji}=1$). Note that $\Phi(t)=\sum_j \pi_j(t)\phi_j(t)$ serves as a normalization factor ensuring conservation of total probability, therefore $\phi_j(t)/\Phi(t)$ identifies a relative fitness.

We build the variational functional with Lagrange multipliers $\lambda_0$ and $\lambda_{ji}(t)$, for the normalization and the flux constraints, respectively:
\begin{widetext}
\begin{eqnarray}
\mathcal{L}[\mathcal{P}'(\Gamma)] = -\sum_{\Gamma} \mathcal{P}'(\Gamma) \log \mathcal{P}'(\Gamma) - \lambda_0\left[\sum_{\Gamma} \mathcal{P}'(\Gamma)-1\right] - \sum_{t,j,i}\lambda_{ji}(t)\left[\sum_{\Gamma} \mathcal{P}'(\Gamma) N_{\Gamma}(j\to i,t)-\pi_j(t) \frac{Q_{ji} \phi_j(t)}{\Phi(t)}\right],
\end{eqnarray}
\end{widetext}
where $\mathcal{P}(\Gamma)=\pi_{i_{0}}\mathcal{P}'(\Gamma)$. This choice simplifies the subsequent calculations without loss of generality.

By taking the functional derivative with respect to $\mathcal{P}'(\Gamma)$ and setting it to zero, we obtain
\begin{equation}
\frac{\delta \mathcal{L}}{\delta \mathcal{P}'(\Gamma)} = -\log \mathcal{P}'(\Gamma) - 1 - \sum_{t,j,i} \lambda_{ji}(t) N_{\Gamma}(j\to i,t) - \lambda_0 = 0,\nonumber
\end{equation}
which leads to
\begin{equation}
\mathcal{P}'(\Gamma) = \frac{1}{Z} \exp \left[\sum_{t,j,i}\lambda_{ji}(t) N_{\Gamma}(j\to i,t)\right],\nonumber
\end{equation}
where $Z = e^{1+\lambda_0}$. Accordingly, the path probability factorizes as
\begin{equation}
\mathcal{P}'(\Gamma)=\prod_{t=0}^{T-1} \frac{\exp \left[\lambda_{j_t,j_{t+1}}(t)\right]}{Z_{j_t}(t)},
\end{equation}
with $Z_{j}(t)=\sum_{i}\exp\left[\lambda_{ji}(t)\right]$. Thus, the elementary transition probability is given by
\begin{equation}
P_{ji}(t)=\frac{\exp\left[\lambda_{ji}(t)\right]}{Z_{j}(t)} = \frac{\exp\left[\lambda_{ji}(t)\right]}{\sum_{k}\exp\left[\lambda_{jk}(t)\right]}.
\end{equation}
The imposed constraint on the mean number of transitions requires that $\pi_j(t) P_{ji}(t) = \pi_j(t) \frac{Q_{ji} \phi_j(t)}{\Phi(t)}$, leading to $P_{ji}(t) = \frac{Q_{ji}\phi_j(t)}{\Phi(t)}$. By comparing this result with the expression previously derived for $P_{ji}(t)$, we obtain the condition
\begin{equation}
\frac{\exp\left[\lambda_{ji}(t)\right]}{\sum_{k}\exp\left[\lambda_{jk}(t)\right]} = \frac{Q_{ji} \phi_j(t)}{\Phi(t)},
\end{equation}
which enforces that the microscopic transition probabilities derived from the variational principle are consistent with the imposed macroscopic constraints on the mean flux, thus connecting the Lagrange multiplier formulation with the relevant biological quantities. The functional form of the transition probability must be $P_{ji}(t) = \frac{Q_{ji}\phi_j(t)}{\Phi(t)}$, to satisfy the flux constraints: this identification is critical in moving from the microscopic  description to the macroscopic one.

In fact, at the macroscopic level, the overall state of the population is described by the frequency vector $\mathbf{x}(t) = \left(x_1(t), x_2(t), \dots, x_{N_{\text{types}}}(t)\right)$ where $x_i(t)$ is the fraction of individuals of type $i$ at time $t$ (with $\sum_{i} x_i(t)=1$). Note that $x_{i}(t)$ can be effectively interpreted as the occupation probability $\pi_i(t)$ of state $i$: accordingly, the evolution of the macroscopic state is given by averaging over the microscopic transitions as
\begin{eqnarray}
x_i(t+1)=\sum_{j=1}^{N_{\text{types}}} x_j(t) P_{ij}(t).\nonumber
\end{eqnarray}
Since the elementary transition probabilities $P_{ji}(t)$ are now known, we finally obtain
\begin{equation}
x_i(t+1)=\sum_{j=1}^{N_{\text{types}}} x_j(t) \frac{Q_{ji} \phi_j(t)}{\Phi(t)},
\end{equation}
which is the discrete replicator-mutator equation.

Therefore, the transition probability $P_{ij}(t)$ provides the building blocks from which the full frequency distribution $\mathbf{x}(t)$ emerges. This two-level description shows how the maximum caliber derivation begins at the microscopic level -- by tracking individual type changes -- and then, through aggregation, yields the macroscopic replicator-mutator dynamics that govern the evolution of the entire population. 

\rev{The above result provides a top-down variational inference from constraints, bridging microscopic dynamics with macroscopic evolutionary behavior, and complementing bottom-up approaches based on probabilistic coarse-graining that connect microscopic stochastic rules (e.g., the Moran process) to their Fokker-Planck limit leading to a replicator-mutator equation~\cite{traulsen2006coevolutionary}.}

\rev{\subsubsection{Extensions and connections to eco-evolutionary statistical mechanics}}

\rev{The variational derivation of the replicator--mutator dynamics presented above can be further extended to encompass a broader class of evolutionary models. In particular, the Crow--Kimura equation (Eq.~(\ref{eq:crow-kimura})) represents a continuous-time generalization of the quasispecies model introduced by Eigen, where selection and mutation jointly act on overlapping generations. 
Remarkably, in the limit of small mutation rates, the equation captures diffusive exploration of the fitness landscape. This analogy has motivated several theoretical developments casting evolutionary dynamics in the language of statistical physics, such as the quantum spin-chain formulation~\cite{saakian2004eigen,saakian2019solution}, where the effective Hamiltonian $\mathcal{H}_{\text{evo}}$ encodes selection and mutation processes, and the mean fitness emerges as the spectral quantity
\begin{eqnarray}
\mathcal{R} = \lim_{\beta \to \infty} \frac{\log \text{Tr}  e^{-\beta \mathcal{H}_{\text{evo}}}}{\beta},
\end{eqnarray}
with $\beta$ an inverse-temperature-like parameter controlling the selectivity of the evolutionary process. In this formulation, the macroscopic quantity $\mathcal{R}$ plays a role which is formally equivalent to that of a free energy, summarizing the asymptotic growth rate of the population. Note that, in the notation of Eq.~(\ref{eq:crow-kimura}), we have $\mathcal{R}\equiv\Phi(\infty)$, as equilibrium is reached in the limit of infinite time.}

\rev{Building on this literature, the Crow--Kimura model has been extended to include ecological feedbacks among phenotypes within a statistical mechanics framework for eco-evolutionary dynamics ~\cite{sireci2024statistical}. Starting from a stochastic process defining the dynamics of the density of individuals with trait $x$ at time $t$ in a given realization as
\begin{eqnarray}
\rho(x,t)=\frac{1}{N}\sum_{i=1}^{N}\delta(x-x_i),\nonumber
\end{eqnarray}
the average over the ensemble of possible realizations is defined by 
\begin{eqnarray}
\varphi(x,t)=\left\langle \frac{1}{N}\sum_{i=1}^{N}\delta(x-x_i)\right\rangle,\quad \int dx~\varphi(x,t) = 1\nonumber
\end{eqnarray}
which in the mean-field (deterministic) approximation leads to the integro-differential equation
\begin{eqnarray}
\partial_t \varphi(x,t) = \int dx' Q_{x'}(x-x') \phi(x',t)\varphi(x',t) - \Phi(t)\varphi(x,t),\nonumber
\end{eqnarray}
where $Q_{x'}(x-x')$ is a mutation kernel describing transitions between phenotypic states, $\phi(x,t)$ is the local fitness field and $\Phi(t)=\int dx \varphi(x,t)\phi(x,t)$ represents the population-averaged marginal fitness. This formalism, grounded in nonequilibrium statistical mechanics, explicitly introduces a partition function over phenotypic configurations and treats $\Phi(t)$ as a free-energy-like potential governing the macroscopic dynamics of the eco-evolutionary ensemble.}

\rev{It is worth noticing that a similar mathematical structure underlies the network density-matrix formalism discussed in Sec.~\ref{sec:hierarchy}. In that framework, considering the network diffusion dynamics governed by the Laplacian operator $\mathbf{L}$, the statistical propagator $e^{-t\mathbf{L}}$ defines the density matrix (see Eq.~(\ref{eq:rho_Z})) and the corresponding generalized free energy $\mathcal{F} = -\frac{1}{t}\log Z$. The analogy becomes explicit by mapping $\mathbf{L}$ onto $\mathcal{H}_{\text{evo}}$ and considering that the scale of fluctuations given by $t$ plays the role of the inverse temperature $\beta$~\cite{de2016spectral,nicolini2018thermodynamics,ghavasieh2020statistical,villegas2023laplacian,ghavasieh2024diversity}, from which it follows that $\mathcal{R} \equiv \mathcal{F}$. Remarkably, this results shows that macroscopic efficiency or growth emerges from microscopic transition operators -- the evolutionary Hamiltonian in sequence space or the Laplacian in network space -- governing diffusive-like exploration of configuration landscapes. From this perspective, one may interpret evolutionary dynamics as a form of diffusion in a latent trait space, where the mutation kernel $Q_{ij}$ or $Q_{x'}(x-x')$ defines the effective connectivity among phenotypes, yielding a ``network of traits'' (as already recognized, in different terms, in the operator formulation of mutation-selection dynamics~\cite{baake2000biological}). Within this latent network, the mean fitness $\mathcal{R}$ quantifies the system's capacity to transmit adaptive information, in the same way that $\mathcal{F}$ quantifies the speed of information flow in the network density-matrix framework. Accordingly, the same thermodynamic formalism that describes hierarchical organization and information transport in complex networks also provides a physically grounded principle for adaptive and evolutionary processes.}


\section{Discussion}\label{sec:discussion}

Let us suppose to capture a proton from a cosmic ray originating from a supernova eight billion light-years away and to compare it to a proton generated in a particle accelerator on Earth, only to find they are indistinguishable. This simple observation highlights the validity of fundamental physical laws across the cosmos. The fundamental question is whether, and under which conditions, the consistency in the physical and the chemical laws have a parallel in biology: would hypothetical fundamental living units discovered in a distant part of our universe resemble those on Earth? Although the exact materials and structures of these units might differ due to unique evolutionary paths and environmental conditions, the fundamental laws of physics and chemistry, consistent across the universe, provide the same foundation for the complexity emerging across many levels of organization, where living systems arise. At those levels, we can safely claim that life is shaped by thermodynamic, informational, and ecological constraints.

As mentioned before, a key concept that characterizes living systems is evolvability~\cite{wagner2013robustness}. 
Accordingly, we should wonder if and how networks within an individual, such as biomolecular and cellular networks, and between individuals, such as ecological interactions, which are governed by nonequilibrium thermodynamics or generalized thermodynamics laws, influence evolvability. 

At the individual level, we have highlighted how living systems need to be complex and adaptive, showing how their behavior on non-evolutionary timescales can be modeled in terms of complex adaptive networks. Fundamentally, living systems behave like thermodynamic engines that transform energy and matter, by using information to extract relevant work that allows them to self-organize, at the expense of their environment. \rev{Remarkably, not all network structures are compatible with adaptation: only specific topologies, such as modular or hierarchically organized networks, can sustain evolvability.} Complex genetic and metabolic circuitries endowed with such topological features provide a trade-off between robustness to perturbations and flexibility to change under selective pressures, creating fertile ground for the emergence of new phenotypes and environmental adaptation~\cite{conrad1990geometry}.
For instance, genetic networks maintain stability while allowing the accumulation of cryptic genetic variation, which can be unlocked under changing environments. Hubs in regulatory networks coordinate gene expression, influencing phenotypic changes and evolvability~\cite{crombach2008evolution,koubkova2018heterologous,helsen2019network}. Metabolic networks enhance adaptability through enzyme promiscuity, enabling organisms to exploit new resources. Even evolvability can be subject to natural selection, favoring mechanisms like modularity and dynamic criticality, which foster adaptation. 

At the ecological level, interactions among organisms drive the evolution of evolvability, adding a layer of complexity beyond the individual level. Co-evolution, \rev{from gene--culture in birds, cetaceans and primates~\cite{whitehead2019reach}, to mutualistic networks~\cite{guimaraes2017indirect}}, continually reshapes adaptive landscapes by introducing indirect effects across species. These indirect interactions, often involving species that are not directly linked, can shape trait evolution: by constantly modifying the adaptive landscape, co-evolution fosters evolvability, enabling species to explore new phenotypic adaptations and respond more effectively to environmental changes~\cite{guimaraes2017indirect}.
Genetic incompatibility, such as that arising from hybridization between species, can also act as a mechanism for increasing evolvability. When a gene hub is replaced by an ortholog from a distantly related species, this can lead to phenotypic diversity: while hybridizations might disrupt key functions, they can also unlock cryptic genetic variation, providing new evolutionary pathways. Therefore, hybridization and genetic incompatibility act as ecological perturbations that drive the exploration of new phenotypic spaces, increasing the overall evolvability of populations~\cite{wagner2008robustness,koubkova2018heterologous}.
Furthermore, horizontal gene transfer (HGT) is another key mechanism through which ecological interactions influence evolvability. Acquiring foreign genes can introduce entirely new functions to an organism, disrupting or enhancing existing genetic networks while creating novel evolutionary opportunities via adaptation to new environments or exploitation of different ecological niches~\cite{wagner2008robustness,koubkova2018heterologous}. Overall, clearly understanding how the intervening complex networks, not limited to regulatory ones~\cite{torres2012criticality,kim2014robustness}, can facilitate adaptive evolution on realistic complex fitness landscapes~\cite{papkou2023rugged} deserves further investigation in the future. \rev{Together, these examples indicate that evolvability is not a property of isolated genomes but an emergent feature of networked interactions across scales.}

To decode the architecture of living systems effectively, one cannot limit to decipher their governing logic: it is necessary to integrate the roles of the intervening complex networks that catalyze, actualize, sustain, and evolve them. Together with the underlying biological and physical principles, we need to focus on the individual and ecological networks that implement and operationalize these principles. In fact, analogous to how thermodynamic machines leverage the laws of thermodynamics to transform energy into usable work, biological networks translate genetic, metabolic and ecological requirements or pressures into functional biological processes. \rev{This raises the question of how such transformations are mechanistically possible.} A powerful metaphor often used to describe the behavior of living systems relies on a parallel with computing machines. While this parallel is useful to describe a variety of features, it is not fully satisfying~\cite{jaeger2024naturalizing}, and it is worth remarking that in empirical settings, once again, the specific functions of a computing machine do not just depend on its logical framework but also on the circuits that execute this logic. 

In the light of the role of circuitry, interactions and interdependency in the major evolutionary transitions considered here, it is plausible to conjecture that living organisms must be complex adaptive systems that co-evolve to maximize their evolvability. 
\rev{These insights suggest a tantalizing, yet to be demonstrated, unifying thermodynamic perspective: if living systems are complex adaptive networks co-evolving under energetic and informational constraints, then evolvability itself might be subject to optimization; a tentative \emph{principle of maximum evolvability}.}
To this aim, the research agenda for the next future is rich. We argue that the current advances in the thermodynamics of systems out of equilibrium, stochastic thermodynamics and physics of complex systems provide a fertile ground for this ambitious research program.

The models discussed herein effectively illustrates this process: within the bounds of specific conditions and constraints, selective forces might effectively drive the development of biological hierarchical modular structures~\cite{hausler2024correlations} that enable an organism to fulfill distinct functions, including adaptation. This yields a vast, yet finite, set of possible network configurations, each uniquely shaped by particular environmental demands, providing a huge potential for evolving evolvability. 

\begin{figure*}[!t]
\centering
\includegraphics[width=\textwidth]{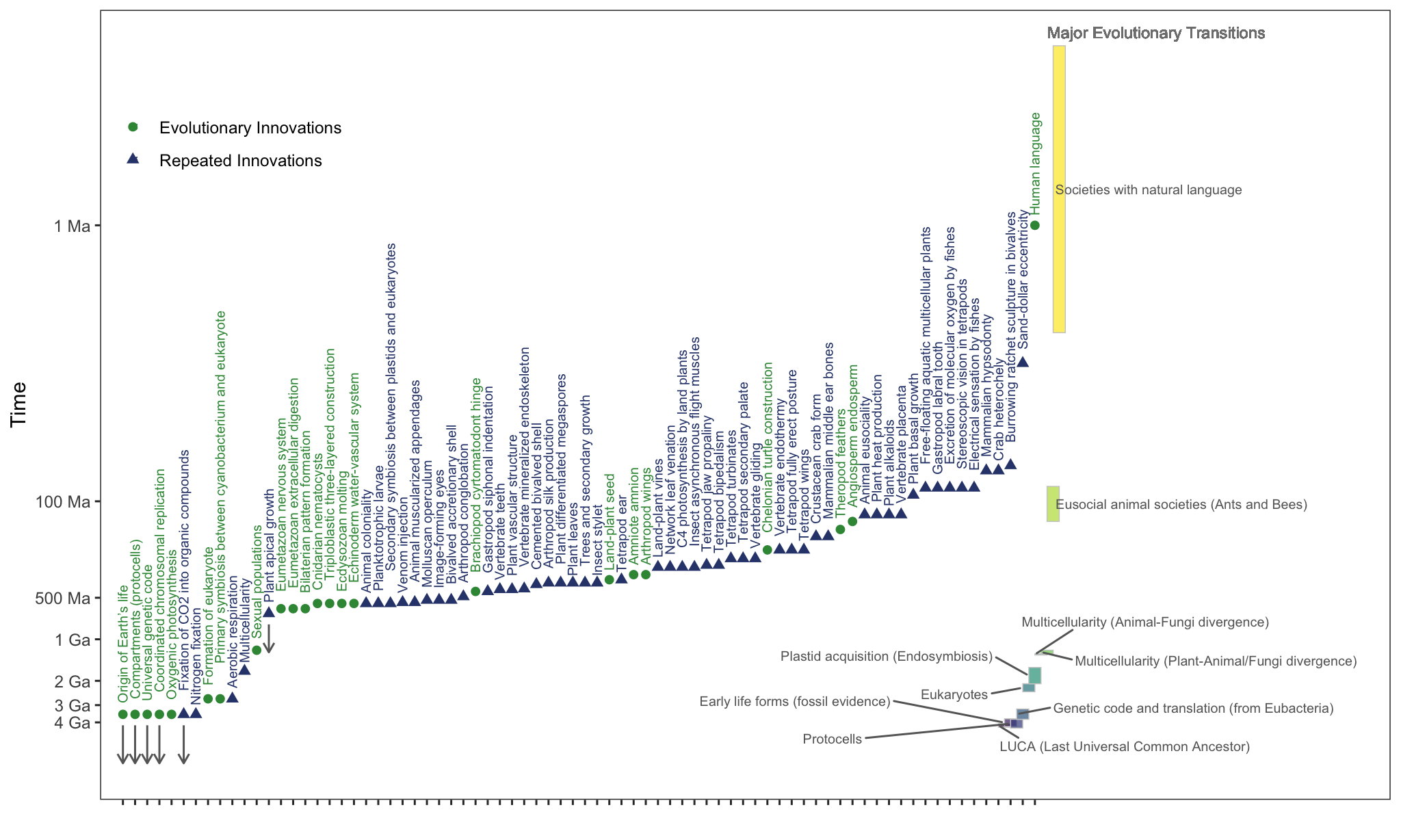}
\caption{\label{fig:evo} \textbf{Mapping major evolutionary transitions and innovations}. Time of appearance of 23 unique evolutionary innovations (filled circles) and 55 innovations (filled triangles) that evolved multiple times (data from~\cite{vermeij2006historical}), suggesting that while early phases of life may have been less replicable, later evolutionary phases show more predictability in terms of functional roles and adaptive changes. On the right-hand side, the filled vertical bars indicate the uncertainty about the major evolutionary transitions on Earth considered in this work, together with other relevant information, from bottom to top: Last Universal Common Ancestor ($[4.09, 4.33]$~Ga)~\cite{moody2024nature}; fossil evidence for early life ($[3.77, 4.28]$~Ga)~\cite{dodd2017evidence}; protocells ($[3.8, 4.4]$~Ga)~\cite{gozen2022protocells}; genetic code and translation (from Eubacteria, $[3.2, 3.8]$~Ga) and the divergence of eukaryotes ($[2.1, 2.4]$~Ga)~\cite{feng1997determining}; acquisition of plastids through endosymbiosis ($[1.6, 2.1]$~Ga)~\cite{strassert2021molecular}; divergence of multicellularity between animals and fungi ($[1.275, 1.3]$~Ga) and between plants and animals/fungi ($[1.2, 1.275]$~Ga)~\cite{feng1997determining}; eusocial animal societies ($[78, 140]$~Ma)~\cite{brady2006evaluating,cardinal2011antiquity}; societies with natural language ($[0.05, 6]$~Ma)~\cite{hillert2015evolving} (and references therein).}
\end{figure*}

This fact suggests that while, on the one hand, a deterministic theory of living systems based on a Newtonian reductionist and deterministic perspectives is unlikely to be achieved~\cite{ulanowicz1999life}, on the other hand there is enough room for moving beyond the Darwinian adaptionist and non-deterministic paradigms~\cite{morowitz1993beginnings}, and build evolutionary theories that are replicable and predictable in functional and ecological terms~\cite{vermeij2006historical,conway2009predictability,conway2010evolution,lassig2017predicting,leigh2022predictable} (see also Fig.~\ref{fig:evo}). Such theories should account for both the logic~\cite{jacob1970logique,sole2024fundamental} and the circuitry that characterize the relational and integrated systems that exhibit self-organization and other emergent phenomena~\cite{rosen1958relational1,rosen1959relational2,ulanowicz1999life}.

\rev{Ref.~\cite{kauffman2023third} asked whether it is possible to prestate all Darwinian preadaptations in the biosphere’s evolution} from 3.7~billion years ago to some point in time in the distant future, such as 400~Myr from now. 
The answer is that this might not be possible, due to inherent limitations in predictability. The question, however, raises a deeper inquiry: what can be predicted, and what cannot? Even if we were to understand the global behavior of life across the universe, with some predictive capacity at large scales, life itself remains a phenomenon localized in space. Therefore, it is worth wondering if life is also localized in time or, in other words, if there are inherent temporal limits to prediction, just as there are spatial ones.

When we consider a system where we have complete knowledge of the underlying laws but uncertainty in the initial conditions, particularly in ecological systems where environmental effects are perfectly modeled, the future evolution of species would still be unpredictable. If the emergence or extinction of species depends on specific environmental states (e.g., the concentration of a substance falling below a certain threshold, or certain environmental conditions being met), we argue that the behavior of living systems cannot be predicted on evolutionary timescales, while it can be on short scales. These dynamics are inherently nonlinear, and most likely chaotic in nature, meaning that beyond a certain temporal scale (i.e., the Lyapunov horizon), it becomes impossible to predict when or if the ideal configuration for a species to appear or disappear will occur, even in a perfectly modeled environment. \rev{Quantifying Lyapunov horizons across biological scales could clarify the temporal limits of evolutionary predictability.}

Meanwhile, some evidence about the predictability of evolution has been reported. Using 25 years of field data about \emph{Timema} stick insects, it has been shown that evolution is more unpredictable when involving multiple selective factors and uncertainty in environmental conditions, rather than when involving nonlinear dynamics (such as feedback loops)~\cite{nosil2018natural}.

Accordingly, while the specific biological networks observed on Earth might be uniquely tailored to our planet's evolutionary trajectory, the underlying statistical organization of these network units likely represents a universal feature of life. This suggests that, across the universe, life may evolve through similar principles of network formation, consistently adapting to diverse environmental challenges while maintaining fundamental organizational patterns. Thus, just as protons are the manifestation of a quantum field and exhibit consistency due to the fundamental physical laws, the basic architectural principles of living systems might also exhibit a universal signature, reflecting a shared architecture for life that maximizes efficiency, adaptability and evolvability.

In the next future, it will be increasingly important to further inspect the interplay between logic and circuitry. We have shown that complex networks with specific configurations are ubiquitous and, \emph{de facto}, unavoidable to achieve specific functions. \rev{As discussed in the Introduction, evolution does not store configurations but generative blueprints. These “life programs” operate across scales -- from molecular to ecological -- encoding the rules for constructing and adapting networks.}
\rev{In this light, Jacob’s tinkerer~\cite{jacob1977evolution}  may be viewed as performing a kind of computation, searching not within configuration space but within the space of life programs.} This point is crucial, since in this case we could view evolution as a search through the space of life programs rather than the space of configurations. In fact, “mutations” in such a life programs space could correspond to faster and more dramatic changes than the ones generated in the configuration space. This perspective aligns with our aim of understanding the underlying processes that enables life's remarkable capacity to evolve: by decoding the architecture of living systems, we begin to uncover those universal principles. \rev{Decoding the architecture of living systems thus means uncovering the universal variational principles that govern their capacity to evolve.}

Nevertheless, while the computational analogy illuminates how life might encode these generative programs, it is crucial not to oversimplify living systems as mere computational devices. If computation may indeed be an important driver of evolution~\cite{hernandez2018algorithmically}, it is also true that biological systems might transcend computational limits as described by classical models~\cite{goni2024biocomputation}. Regardless of this debate, our central thesis remains: evolution shapes and refines the life programs that govern the architecture and adaptability of life, and we are just starting to understand the logic~\cite{benenson2012biomolecular,sole2024fundamental}. 

Identifying the fundamental mechanisms, such as generalized variational principles (e.g., the minimization of a suitable action or the maximization of a thermodynamic observable, such as the entropy production~\cite{lotka1922contribution,lotka1922natural,martyushev2006maximum} or the caliber~\cite{presse2013principles}), to describe the evolvability of living systems in terms of the interplay between their underlying logics and corresponding circuitries remains an unsolved problem and an exciting, yet challenging, research direction for the near future.

\heading{Acknowledgments} The author thanks O. Artime, A. Beretta, D. Brockmann, P.L. Cerni, V. d'Andrea, E. Frey, J. Grilli, T. Gross, H. Jensen, A. Kolchinsky, J.P. Hall, M.A. Mu\~{n}oz, J. Nauta, L. Pessoa, W. Ratcliff, R. Sol\'e, S. Suweis, D. Zanchetta and H. Zenil for fruitful and interesting discussions. This work has been partially supported by the Human Frontier Science Program Organization (HFSP Ref. RGY0064/2022), the EU's Horizon Europe research and innovation programme under grant agreement No. 101186013 (VirHox), and the INFN grant “LINCOLN”.

\bibliographystyle{apsrev4-1}
\bibliography{biblio}

\end{document}